\documentclass[twocolumn,secnumarabic,amssymb, nobibnotes, aps, prd]{revtex4-1}

\usepackage[latin1]{inputenc}
\usepackage[T1]{fontenc}
\usepackage[english]{babel}
\usepackage[dvips]{graphicx}
\usepackage{enumerate,amsthm,amsmath,amssymb,color,graphicx,bbm,float,framed}
\usepackage{natbib}

\newtheorem{theo}{Theorem}

\begin{document}

\title{Security of Continuous-Variable Quantum Key Distribution with Discrete Modulation against General Attacks}%
\author{Jian Zhou$^{1}$, Ying Guo$^{1,2,*}$, Duan Huang$^{1}$ and Guihua Zeng$^2$}%
\affiliation{$^1$School of Information Science and Engineering,
Central South University, Changsha {\rm 410083}, China \\
$^2$State Key Laboratory of Advanced Optical Communication Systems and Networks,
Department of Electronic Engineering, Shanghai Jiao Tong University, Shanghai {\rm 200240}, PR China.}
\email[Corresponding author:]{yingguo@csu.edu.cn}
\date{\today}%
\begin{abstract}
We provide a security analysis of continuous-variable quantum key distribution (CVQKD)
with discrete modulation against general attacks in a realistic finite-size regime.
To realize this goal, we first prove security of the continuous-variable quantum key distribution protocol
with discrete modulation against collective attacks by using the reliable tomography of the covariance matrix, leading to the reliable and tight error bounds in the derived confidence regions.
Combining the proof with de Finetti reduction, the  discrete-modulation-based continuous-variable quantum key distribution (DM-CVQKD) is proved to
be secure even exposing to general attacks. Specially, we use an energy test to truncate the Hilbert space globally to provide security.
\end{abstract}
\maketitle

Quantum key distribution (QKD) enables two distant parties, Alice and Bob, who have access to
an authenticated classical channel to share secret keys even in the presence of an eavesdropper~\cite{RevModPhys.81.1301}.
The traditional QKD protocols can be divided into two families, the discrete-variable (DV) QKD~\cite{PhysRevLett.94.230504}
that based on photon counting techniques and continuous-variable (CV) QKD which relying on
detections. CVQKD, which has stable reliable light resource and high detection efficiency, may be practically compatible with classical optical communication over DVQKD~\cite{PhysRevLett.91.057901}. Unfortunately, despite all these advantages, CVQKD is still not considered
as a true alternative to DVQKD because of the short secure distance~\cite{PhysRevA.90.042329}. Different from DVQKD, the task of classical post-processing for Alice and Bob
extracting a key from continuous random values is more complicated and the post-processing efficiency is also achieving its limit~\cite{madsen2012continuous}. For example, an initial
CVQKD  protocol using discrete modulation (DM) was proposed to
realize the task of distributing secret keys over long distance as well as simplifying both
the modulation scheme and the key extraction process~\cite{PhysRevLett.102.180504,PhysRevA.86.022338}.

Currently, proving the security of QKD against general attacks in a realistic finite-size regime has become an open problem in the theoretical quantum cryptography~\cite{PhysRevLett.102.020504}.
For DVQKD protocol, composable security~\cite{muellerquade2010composability} was established for some classical protocols
i.e. BB84 protocol~\cite{Bennett1984Quantum}. The finite-key analyses based on the uncertainty relation for smooth entropies~\cite{Tomamichel2012Tight}
and the large deviation theory~\cite{Curty2012Finite} were also suggested.
As for CVQKD counterpart, a composable security proof~\cite{PhysRevLett.109.100502,PhysRevA.90.042325,Gehring2015Implementation} of squeezed states was achieved
from an entropic uncertainty principle~\cite{Furrer2014Position}. Recently,
the composable security proof of CVQKD with coherent states has been demonstrated~\cite{PhysRevLett.114.070501}.
Based on Gaussian de Finetti reduction, which exploits the invariance of CV protocols
under the action of unitary group $U(n)$ and the generalized $SU(2,2)$ coherent states,
the security of CVQKD with coherent states against general attacks was proved~\cite{PhysRevLett.118.200501}.
Up to now, the finite-size effects of the DM-CVQKD protocol have only been considered on the parameter estimation (PE) procedure~\cite{PhysRevA.81.062343}, however the security against general attacks is still far from being completely understood.

In this work, we present the security proof of the DM-CVQKD with reverse reconciliation (RR) valid against general attacks.
The contributions of this work are twofold. In contrast to existing finite-size effects on PE of the DM-CVQKD, we provide, for the first time,
the composable security proof against collective attacks in finite-key regime. Moreover, we apply the de Finetti reduction to obtain the security
against general attacks on the basis of the former. The first procedure involves the reliable tomography
of the covariance matrix and error correction (EC), and the latter exploits the invariance of CV
protocols under the actions of the unitary group, leading to the generalized coherent states. Besides,
truncating the total Hilbert space to a finite-dimensional one with an energy test is also necessary to implement de Finetti reduction.
Since the four-state protocol outperforms other DM-CVQKD protocols, we consider the case of the former in what follows.

\textit{Security Definition}.---In the DM-CVQKD protocol, the state can be expressed in Fock state~\cite{PhysRevA.83.042312}. We take
four-state modulation for example,
\begin{align}
\begin{array}{cccc}
|\phi_k\rangle=\frac{e^{-\frac{|\alpha|^2}{2}}}{\sqrt{\lambda_k}}\sum^\infty_{n=0}\frac{\alpha^{4n+4}}{\sqrt{(4n+k)!}}|4n+k\rangle.
\end{array}
\end{align}
The resulting DM-CVQKD system is a canonical infinite dimensional quantum system
composed of an ensemble of $4\mathcal{N}$ modes described by a Hilbert space.

To prove the security of the protocol, we follow the definition in ref.~\cite{mullerquade2009composability}. A list of properties \textit{ideal} we expect the protocol to achieve is first given, and then the security of the \textit{real} protocol is defined by indistinguishability from the \textit{ideal} case. Normally, we call these properties \textit{secrecy}, \textit{correctness} and \textit{robustness}. The protocol will be considered
as \textit{robust} if it outputs nontrivial keys when the adversary is passive which means the adversary does not disturb quantum communications. We consider the strongest type of security by exposing the protocol to \emph{general attacks}, where an adversary may tamper with the arbitrary signals exchanged between Alice and Bob over quantum channels~\cite{cai2009finite-key}. The classical channel can be
monitored by the adversary. In this work, we first prove the security of the DM-CVQKD protocol against collective, and then reduce the \textit{general attacks} to the collective one with de Finetti
reduction.

A protocol is called \textit{correct}, if the outputs of the protocol on Alice and Bob's side are identical. It is called $\epsilon_\mathrm{cor}$-correct if it is $\epsilon_\mathrm{cor}$-indistinguishable from a correct protocol. Particularly, we have $\mathrm{Pr}[S_A \ne S_B] \leq \epsilon_{\mathrm{cor}}$~\cite{Tomamichel2012Tight}. A protocol is called \emph{secrecy} if the protocol produces a key $S$ which is uniformly distributed and
independent of the states of the system \textit{E} held by the adversary. A key is called $\epsilon$-secret from \textit{E} if it is $\epsilon$-close to a uniformly distributed key that is uncorrelated with the
adversary~\cite{konig2007small}, that is
\begin{align}
\min_{\rho_E}{\frac{1}{2}\|\rho_{SE} -\omega_{S}\otimes\rho_{E}\|_1}\leq\epsilon,
\end{align}
where $\omega_{S}$ is the completely mixed states on $S$. A  protocol is called secret, if, for any attacks strategy, $\epsilon=0$ whenever the protocol outputs a key.
It is called $\epsilon_\mathrm{sec}$-secret, if its output key is $\epsilon$-secret with $(1-p_\mathrm{abort})\epsilon\leq\epsilon_\mathrm{sec}$, where $p_\mathrm{abort}$ is the probability that the protocol aborts.
This probability depends on the strategy of the adversary, which is on the input state $\Psi_{ABE}$. Indeed, the adversary can always choose to cut the line between Alice and Bob and hence make sure that the protocol always aborts. This is fine since the key will nevertheless be secure. Another important parameter is the \textit{robustness}, $\epsilon_{\mathrm{rob}}$,  which corresponds to the abort probability when the adversary is passive, and if the characteristics of quantum channels are conform to what was expected. For instance, in the case of the DM-CVQKD, a typical quantum channel can be characterized by its transmittance $T$ and excess noise $\xi$. A protocol is $\epsilon$-secure if it is $\epsilon_{\mathrm{sec}}$-secret and $\epsilon_{\mathrm{cor}}$-correct with $\epsilon_{\mathrm{sec}} + \epsilon_{\mathrm{cor}} \leq \epsilon$.

An operational way of quantifying the security is by bounding the diamond distance between two completely positive trace-preserving (CPTP) maps $\mathcal{E}$ and $\mathcal{F}$, where $\mathcal{E}$ denotes the practical DM-CVQKD protocol and $\mathcal{F}$ is an ideal version of the same protocol. The ideal version can be achieved by concatenating $\mathcal{E}$ and a (virtual) protocol $\mathcal{P}$ that replaces the final keys $S_A$ and $S_B$ by a perfect key $S$.
We define $\mathcal{F}=\mathcal{P}\circ\mathcal{E}$ and the protocol $\mathcal{E}$ is $\epsilon$-secure if $\frac{1}{2}\|\mathcal{E}-\mathcal{F}\|_\diamond\leq\epsilon$, which will be detailedly discussed in the de Finetti reduction part.

\textit{Protocol Definition}.---As shown in Table \ref{table1}, Alice prepares a string four-dimensional vectors, which are picked from one
of the three distributions: discrete modulation states ($2n$ random variables used for key distillation), decoy states ($2m$ make the mixture of these states indistinguishable from a Gaussian state)
and Gaussian states ($2k$ used for parameter estimation). Alice sends these states through a linear quantum channel to Bob, who performs
heterodyne detection. Alice and Bob use a reverse reconciliation and the sign of measurement result to encode the raw key bit.
Bob also sends some side information to Alice (syndrome of $U$ for a error correcting code $C$ agreed on in advance), and Alice compares its own gauss $\hat{U}$
with Bob's string. The comparison results in whether the protocol is kept or aborted. Alice and Bob divide Gaussian state $X$ and $Y$ into two parts,
$\{X_1,X_2\}$ and $\{Y_1,Y_2\}$, respectively. Then they use $X_1$ and $Y_1$ to give the confidence region for $X_2$ and $Y_2$, or vice versa. The key will not be
kept except the data falls within the confidence interval. Finally, Alice and Bob adopt privacy amplification to increase the secrecy of the raw key.

\begin{table}
\caption{\label{table1} The DM-CV QKD
protocol.}
\begin{ruledtabular}
\begin{tabular}{lp{2in}}
\textbf{State Preparation:}---Alice chooses one of the three distri-\\
butions: discrete modulation states, decoy states (discarded \\
at the end of the protocol) and Gaussian states each time.  \\
Alice randomly draws a random orthogonal transformation $R$ \\
from the orthogonal group and computes vector $x'=Rx$ wh- \\
ich will be used for state modulation. \\
\textbf{Distribution:}---Alice sends these states to Bob over a linear \\
channels, which can be modeled by a transmission $T$ and a \\
excess noise $\xi$. \\
\textbf{Measurement:}---Bob measure their modes with heterodyne \\
detection and obtains a vector $y'$. Alice describes $R$ to Bob \\
and Bob applies $R'$ to his vector $y'$ and obtains $y=R^{-1}y'$.\\
\textbf{Data Classification:}---Alice reveals which data should be \\
kept for key distillation, which one should be discarded  \\
(decoy states) and which one should be used for parameter \\
estimation (Gaussian states). \\
\textbf{Error Correction:}---Bob sends side information to Alice \\
over classical channels, typically the syndrome $U$ of his string \\
relative to a binary code they agree on beforehand. Alice out- \\
puts a guess $U_A$ for the string of Bob. Bob computes a hash \\
of $U$ with length $\lceil\log_{1/\epsilon_{\mathrm{cor}}}\rceil$ and sends it to Alice compares it \\
with its own hash. If both hashes coincide, the protocol resu- \\
mes, otherwise it aborts. The value $\mathrm{leak}_{\mathrm{EC}}$ corresponds to the \\
total number of bits sent by Bob during the EC phase. \\
\textbf{Parameter Estimation:}---Bob sends $n_{\mathrm{PE}}$ bits of information \\
to Alice that allow her to calculate $\{\gamma_a, \gamma_b,\gamma_c\}$. If $\gamma_a \leq \Sigma_a^{\max}$, \\
$\gamma_b \leq \Sigma_b^{\max}$ and $\gamma_c \geq \Sigma_c^{\min}$, then the protocol continues. Other- \\
wise it will be aborted. \\
\textbf{Privacy Amplification:}---Alice and Bob apply a random \\
universal$_2$ hash function to their respective strings, resulting \\
in two strings $S_A$ and $S_B$ of size $l$.\\
\end{tabular}
\end{ruledtabular}
\end{table}

\textit{Analysis for Collective Attacks}.---We present the main results, quantifying the security of the protocol $\mathcal{E}$ against Gaussian collective attacks. The output of the DM-CVQKD protocol is
nontrivial when the adversary is passive. Besides, It is both $\epsilon_\mathrm{cor}$-correct and $\epsilon_\mathrm{sec}$-secret
against collective attacks, given that the length $l$ of the secret $S_A$ is selected approximately for a given set of observed values.
The correctness of the protocol is guaranteed by its error correction step. The hash $U$ used for EC has a length of $\lceil\log_2{\frac{1}{\epsilon_\mathrm{cor}}}\rceil$,
which restricts the error probability $\mathrm{Pr}[S_A\neq S_B]\leq\epsilon_\mathrm{cor}$. The PE procedure gives the confidence regions of
the covariance matrix.

\begin{theo}
The protocol is $\epsilon$-secret against collective attacks
if $\epsilon=\epsilon_\mathrm{PE}+\epsilon_\mathrm{cor}+\epsilon_\mathrm{ent}+\epsilon_\mathrm{sm}$ and the length $l$ of bit string $S$, which forms the secret key $S_A$, satisfies the constraint
\begin{align}
\begin{array}{cc}
 l&\leq 2n[2\hat{H}_{\mathrm{MLE}}(U)-f(\Sigma^{\mathrm{max}}_a,\Sigma^{\mathrm{max}}_b,\Sigma^{\mathrm{min}}_c)] \\
 &-\mathrm{leak}_\mathrm{EC}-\Delta_{\mathrm{AEP}}-\Delta_{\mathrm{ent}},
 \end{array}
\end{align} where $\hat{H}_\mathrm{MLE}(U)$ denotes the empiric entropy of $U$, $\mathrm{leak}_{EC}=\lceil\log_2{\frac{1}{\epsilon_\mathrm{cor}}}\rceil$ corresponds to the total number of bits sent by Bob during the error correction phase, $\Delta_{\mathrm{AEP}}:=\sqrt{n}(16+\log\frac{2}{\epsilon^2}+8\sqrt{\log\frac{2}{\epsilon^2}})+4\frac{\epsilon}{p}+\log\frac{2}{p^2}$, $\Delta_{\mathrm{ent}}:=n\log n\sqrt{2\log\frac{2}{\epsilon_\mathrm{ent}}}$ and $f$ is the function computing the Holevo information between Eve and Bob's measurement result for a Gaussian state with covariance matrix parametrized by $\Sigma^{\mathrm{max}}_a$, $\Sigma^{\mathrm{max}}_b$ and $\Sigma^{\mathrm{min}}_c$.
\end{theo}

In security analysis of collective attacks, we consider parameter estimation (PE).
As explained previously, the aims of the PE processing are to give a confidence region of the covariance matrix and sift out
the data the pass the test. We follow the quantum state tomography in Ref.~\cite{PhysRevLett.109.120403} and
take quantum tomography as a process symmetrizing the $2k$ systems and making a prediction of the remaining $k$ systems with the measurement results of the first $k$ systems.
The balance between failure probability of prediction and size of the confidence interval need to be kept while smaller confidence interval means larger error probability but higher secret key rate.

A problem concerning the tomography of DM system comes up that the covariance matrix is priori unbounded. The solution to this problem lies in approximately symmetrizing the state
$\rho^{2k}$ before measuring $k$ subsystems and inferring properties for the remaining $k$ systems. We briefly introduce the tomography of the input state $\rho^{2k}$
of the DM-CVQKD protocol in three steps:
\begin{enumerate}[{(1)}]
\item \textit{Active symmetrization}.---Choose a subset $\mathcal{F}$ of the orthogonal group and for Alice and Bob,  apply the same element $f\in\mathcal{F}$ to their $2(n+m+k)$ modes,
outputting a new state $\tilde{\rho}^{2(n+m+k)}$. This operation can also be implemented by applying the same element to their data before postprocessing.
\item \textit{Distribution and measurement}.---Alice and Bob classify the states to three types and use the Gaussian states for tomography.
They distribute $\tilde{\rho}^k_1$ to $A_1$ and $B_1$. Similarly, the remaining states $\tilde{\rho}^k_2$ are given to $A_2$ and $B_2$.
The parties measure their own data with heterodyne detection. $A_1$ and $B_1$ yield $X_1$ and $Y_1$ while $A_2$ and $B_2$ obtain $X_2$ and $Y_2$, where $X_1,Y_1,X_2,Y_2\in\mathbb{R}^{2n}$.
\item \textit{Parameter estimation}.---$B_1$ sends some information to $A_1$ through side channel to help her learn the $\|X_1\|^2$, $\|Y_1\|^2$ and $\langle X_1,Y_1\rangle$. After that $A_1$ will derive a confidence interval for $\tilde{\rho}_2$. Analogously, $A_2$ could also compute a confidence interval for that of $\tilde{\rho}_1$.
\end{enumerate}

Based on $\|X\|^2$, $\|X\|^2$ and $\langle X,Y\rangle$, Alice computes $\gamma_a$, $\gamma_b$ and $\gamma_c$ as follows (see Ref.~\cite{Supplementalmatrial}):
\begin{align}
&\gamma_a:=\frac{1}{2k}[1+3\sqrt{\frac{\log(36/\epsilon_\mathrm{PE})}{k}}]\|X\|^2-1,\\
&\gamma_b:=\frac{1}{2k}[1+3\sqrt{\frac{\log(36/\epsilon_\mathrm{PE})}{k}}]\|Y\|^2-1,\\
&\gamma_c:=\frac{1}{2k}\langle X,Y\rangle-6\sqrt{\frac{\log(144/\epsilon_\mathrm{PE})}{k^3}}(\|X\|^2+\|Y\|^2).
\end{align}
The probability that the PE test passes, but Eve's information computed with the covariance matrix
characterized by $\{\Sigma^\mathrm{max}_a,\Sigma^\mathrm{max}_b,\Sigma^\mathrm{min}_c\}$ is underestimated, is upper bounded by $\epsilon_\mathrm{PE}$.
The parameter estimation is designed as $[\gamma_a\leq\Sigma^\mathrm{max}_a]\wedge[\gamma_b\leq\Sigma^\mathrm{max}_b]\wedge[\gamma_c\geq\Sigma^\mathrm{min}_c]$ and
Eve's information is upper bounded by $f(\Sigma_a,\Sigma_b,\Sigma_c):=g(\nu_1)+g(\nu_2)-g(\nu_3)$, where $\nu_1$ and $\nu_2$ are the symplectic eigenvalues of the covariance matrix
\begin{align}
\left[
\begin{matrix}
\Sigma^\mathrm{max}_a I_2 & \Sigma^\mathrm{min}_c \sigma_z  \\
\Sigma^\mathrm{min}_c\sigma_z  &  \Sigma^\mathrm{max}_b I_2 \\
\end{matrix}
\right].
\end{align}
$\nu_3=\Sigma^\mathrm{max}_a-(\Sigma^\mathrm{min}_c)^2/(1+\Sigma^\mathrm{max}_b)$, $\sigma_z=\mathrm{diag}(1,-1)$, and $g(x):=\frac{x+1}{2}\log_2{\frac{x+1}{2}}-\frac{x-1}{2}\log_2{\frac{x-1}{2}}$.

In order to prove the security of the DM-CVQKD protocol, we restrict to collective attacks as the general attacks can be reduced to Gaussian collective
attacks via post-selection technique or de Finetti reduction. Then, the asymptotic equipartition property (AEP) allows one to compute the smooth
min-entropy as a function of the von Neumann entropy of a single subsystem. We first introduce the leftover hash lemma and then use the AEP to calculate the smooth min-entropy.
The $Leftover$ $Hash$ $Lemma$ provides that the extractable bits, $l$, is slightly smaller than the min-entropy of $A$ conditioned on $B$, $H_\mathrm{min}(A|B)$~\cite{tomamichel2011leftover}, i.e.,
$
l\leq H_{\mathrm{min}}(A|B).
$
This claim requires the data satisfying perfect uniform randomness. The smooth min-entropy is introduced to extend this concept to a case of approximately uniform randomness.
In the four-state protocol, the $\epsilon$-smooth min-entropy of $A^n$ conditioned on $B^n$ of state $\tau_{AB}$, $H^{\epsilon}_\mathrm{min}(A^n|B^n)_\tau$, is bounded by Shannon entropy
of $A^n$ conditioned on $B^n$ , $H(A^n|B^n)_\tau$. Namely, we have
\begin{align}
H^\epsilon_{\mathrm{min}}(A^n|B^n)_{\tau}\geq H(A^n|B^n)_{\tau}-\Delta_{AEP},
\end{align}
with
\begin{align}
\begin{array}{cc}
\Delta_\mathrm{AEP}&=\sqrt{n}(16+\log\frac{2}{\epsilon^2}+8\sqrt{\log\frac{2}{\epsilon^2}})\\
&+4\frac{\epsilon}{p}+\log\frac{2}{p^2}.
\end{array}
\end{align}
Because of the non-zero failure probability in error correction (EC), the conditional state is no longer guaranteed to be a tensor-power.
Indeed, the conditioned state has the form
\begin{eqnarray}
\tau^n=p^{-1}\Pi\rho^{\otimes n}\Pi,\nonumber
\end{eqnarray}
where $\Pi$ is a projector operator and $p=\mathrm{Tr}(\Pi\rho^{\otimes n}\Pi)$ is the success probability of EC.

In DM-CVQKD, the random variables satisfy the independent and identical distribution and have the same probability to fall to $n$ different quadrants, where $n$ corresponding to the number of modulation states.
Unfortunately, the practical entropy $H(U)=-\sum^4_{i=1}p_i\log{p_i}$ is lower than that of the ideal one when considering the finite case and can be bounded by the maximum likelihood estimator of $H(U)$
base on Antos and Kontoyannis's theorem~\cite{antos2001convergence}:
\begin{align}
nH(U)\geq n\hat{H}_\mathrm{MLE}-\Delta_\mathrm{ent},
\end{align}
where $\Delta_\mathrm{ent}:=n\log n\sqrt{2\log\frac{2}{\epsilon_\mathrm{ent}}}$.

\textit{Analysis for General attacks}.---Security of the DM-CVQKD protocol for the general attacks can be reduced to the case of collective attacks under certain conditions.
The DM protocol satisfies the experimental verifiable conditions that the de Finetti theorem holds for states of infinite-dimensional systems~\cite{renner2009de}.
The first property concerns the active symmetrization which requires the protocol being covariant under the action of the unitary group.
We apply an active symmetrization step to the state $\rho_{AB}$ before applying the protocol, such an invariance can also be enforced by symmetrizing the classical data held by Alice and Bob. For the second property,
the key is required to be computed by two-universal hashing in the privacy amplification step. This criterion is not restrictive because two-universal
hashing is optimal with respect to the extractable length. In addition, the dimension of the relevant subspace $\overline{\mathcal{H}}$ of the signal space $\mathcal{H}$ in DM protocol
is small, compared with $N$, which is required by property 3.

The de Finetti representation theorem allows us to drop this assumption, implying that security holds against general attacks.
A protocol is secure against general attacks only when the map $\mathcal{E}$ approximately equals to an ideal protocol $\mathcal{F}$.
The protocol is said to be $\epsilon$-secure if the diamond distance between the two maps is less than $\epsilon$: $\|\mathcal{E}-\mathcal{F}\|_\diamond\leq\epsilon$.
Compared with security analysis of the DV protocols~\cite{sheridan2010finite-key,PhysRevA.82.030301}, an important technicality is that we need to truncate
the total Hilbert space globally to replace it by a finite-dimensional one. We use an energy test $\mathcal{T}$ to bound the number of measured modes below some threshold.
The security of the protocol $\mathcal{E}$ is then reduced to two parts~\cite{PhysRevLett.110.030502}:
\begin{align}
\|\mathcal{E}-\mathcal{F}\|_\diamond\leq\|\overline{\mathcal{E}}-\overline{\mathcal{F}}\|_\diamond+2\|(\mathbb{I}-\mathcal{P})\circ\mathcal{T}\|_\diamond.
\end{align}
where we used the fact that the CP maps $\mathcal{E}_0$ and $\mathcal{F}_0=\mathcal{S}\circ\mathcal{E}_0$ cannot increase the diamond norm.
Alice and Bob perform a rotation operation on their $s+t$ modes and measure the last $t$ modes of their data with heterodyne detection.
The energy test passes if the average energy per mode is below $d_A$ for Alice and $d_B$ for Bob. The thresholds $d_A$ and $d_B$ should be chosen
properly to ensure that the energy test passes with large success probability.

Since the coherent states in DM protocol in Hilbert space displays the permutation invariance and rotation symmetry, the generalized coherent state is given by~\cite{Leverrier2016}
\begin{equation}
|\Lambda,n\rangle=\det(1-\!\Lambda\Lambda^\dag)^{ \frac{n}{2}}\exp{(\mathop{\sum}^2_{i,j=1}\lambda_{ij}Z_{ij})}|\mathrm{vacuum}\rangle,
\end{equation} which resolve the identity on the symmetric subspace~\cite{PhysRevLett.118.200501}:
\begin{align}
\int_D|\Lambda,n\rangle\langle\Lambda,n|d\nu_n(\Lambda)=\mathbb{I}_{F^{U(n)}_{2,2,n}}.
\end{align}
In order to obtain an operator with finite norm, we consider the finite-dimensional subspace $F^{U(n)\leq K}_{2,2,n}$ by bounding the maximum
energy of the states. In Ref.~\cite{PhysRevLett.118.200501}, an approximate resolution of the identity is given by
\begin{eqnarray}
\int_{\mathcal{D}_\eta}|\Lambda,n\rangle\langle\Lambda,n|d\nu_n(\Lambda)\geq(1-\epsilon)\Pi_{\leq K},
\end{eqnarray}
when  the coherent states are restricted to
$\Lambda\in\mathcal{D}_\eta$ with $\mathcal{D}_\eta=\{\Lambda\in\mathcal{D}:\eta\mathbb{I}_2-\Lambda\Lambda^\dag\geq0\}$ for $\eta\in[0,1]$.

The finite energy version of the de Finetti theorem allows us to bound the diamond norm of maps, provided that the total number of the input state is upper bounded by $K$.
If $\mathcal{E}_0$ is $\epsilon$-secure against collective attacks, then
\begin{eqnarray}
\|\overline{\mathcal{E}}-\overline{\mathcal{F}}\|_\diamond=\|\mathcal{R}\circ(\mathcal{E}_0-\mathcal{F}_0)\circ\mathcal{P}^{\leq K}\|_\diamond\leq2T(n,\eta)\epsilon, \nonumber
\end{eqnarray}
where $K=\max\{1,n(d_A+d_B)\frac{1+2\sqrt{\frac{\ln\frac{8}{\epsilon}}{2n}}+\frac{\ln\frac{8}{\epsilon}}{n}}{1-2\sqrt{\frac{\ln(8/\epsilon)}{2k}}}\}$ and $T(n,\eta)=\frac{K^4}{12}$.

\begin{theo}
If the DM protocol $\mathcal{E}_0$ is $\epsilon$-secure against Gaussian collective attacks,
then the protocol $\mathcal{E}=\mathcal{R}\circ\mathcal{E}_0\circ\mathcal{T}$ is $\varepsilon$-secure
against general attacks with
\begin{align}
\varepsilon=(2+{K^4}/{6})\epsilon.
\end{align}
where $\mathcal{R}$ is privacy amplification and $\mathcal{T}$ is the energy test.
\end{theo}
We find the map  $\mathcal{R}$ reduces the final key of the protocol by $\lceil2\log_{2}\binom{K+4}{4}\rceil$.
This result provides security against general attacks with a prefactor $(2+{K^4}/{6})$ based on the security against collective attack
derived in previous part.

\textit{Discussion}.---We have proved the security of the DM-CVQKD protocol against general attacks in the finite regime.
This target is achieved with two steps. It is first proved to be secure against Gaussian collective attacks, and then is extended to the general attacks by using Gaussian de Finetti reduction.
This proof exploits the rotation symmetrization and active symmetrization operations which are crucial for the security analysis. In the PE process,
we use quantum state tomography to derive confidence intervals for the covariance matrix involving the parameters $\{\Sigma^{\mathrm{max}}_a,\Sigma^{\mathrm{max}}_b,\Sigma^{\mathrm{min}}_c\}$.
The {\it leftover hash lemma} and AEP are also used for bounding shannon entropy of the DM protocol.
We use an energy test to ensure that the state shared between Alice and Bob is suitably described by assigning the low-dimensional Hilbert space to each mode.
The bound provided by security against collective attacks suggests the security of DM protocol against general attack via the Gaussian de Finetti reduction.

We thank Cosmo Lupo, who provided very useful comments on a preliminary version of this manuscript.

\bibliographystyle{apsrev4-1}
\bibliography{reference}
\end{document}


\title{Supplementary material:`` Security of Continuous-Variable Quantum Key Distribution with Discrete Modulation against General Attacks"}%
\author{Jian Zhou$^{1}$, Ying Guo$^{1,2,*}$, Duan Huang$^{1}$, and Guihua Zeng$^{2}$}%
\affiliation{$^1$School of Information Science and Engineering,
Central South University, Changsha {\rm 410083}, China \\
$^2$State Key Laboratory of Advanced Optical Communication Systems and Networks,
Department of Electronic Engineering, Shanghai Jiao Tong University, Shanghai {\rm 200240}, PR China.}
\email[Corresponding author:  ]{yingguo@csu.edu.cn}
\date{\today}%
\maketitle

In this appendix, we first recall definitions related to the composable security of quantum key distribution (QKD) with discrete-modulation (DM) under collective attacks in Section I. In Section II, we give a complete description of the QKD protocol $\mathcal{E}_0$ that is secure against collective attacks. In Section III, we show the parameter estimation (PE) procedure and establish an upper bound for the failure probability of the PE test.  In Section IV, we provide the smooth min-entropy of a conditional state. In Section V, the security analysis of DM-CVQKD protocol against general attacks is reduced to that of Gaussian collective attacks when the photon number of the input states is bounded.

\section{Quantum Key Distribution and Composable Security}

In the continuous variable (CV) quantum cryptography, coherent state can be expanded in Fock state,
\begin{align}
\begin{array}{cccc}
|\alpha\rangle=e^{-\frac{|\alpha|^2}{2}}\sum^\infty_{n=0}\frac{\alpha^n}{\sqrt{n!}}|n\rangle.
\end{array}
\end{align}
Analogously, the state in discrete modulation (DM) can be expressed in Fock state. Without loss of generality, we take
four states as an example,
\begin{align}
\begin{array}{cccc}
|\phi_k\rangle=\frac{e^{-\frac{|\alpha|^2}{2}}}{\sqrt{\lambda_k}}\sum^\infty_{n=0}\frac{\alpha^{4n+4}}{\sqrt{(4n+k)!}}(-1)^n|4n+k\rangle,
\end{array}
\end{align}
where $\lambda_{0,2}=\frac{1}{2}e^{-\alpha^2}[\cosh(\alpha^2)\pm\cos(\alpha^2)]$, $\lambda_{1,3}=\frac{1}{2}e^{-\alpha^2}[\sinh(\alpha^2)\pm\sin(\alpha^2)]$ for $k\in\{0,1,2,3\}$. A CV system with DM is a canonical infinite dimensional quantum system
composed of an ensemble of $4\mathcal{N}$ modes described by a Hilbert space, $\mathcal{H}$, corresponding to the four $n$-mode Fock spaces: $\mathcal{H} \cong (F_0\bigoplus F_1\cdots \bigoplus F_3)^{\otimes n}$ where $F_i = \mathrm{Span}(|i\rangle, \ldots, |4k+i\rangle, \ldots)$ and $|k\rangle$ is a $k$-photon Fock state. For simplicity, we denote $F=F_0\bigoplus F_1\cdots \bigoplus F_3$. The fact that this space is infinite-dimensional makes the security analysis of the protocol much more involved than that of BB84 protocol, already for the case of collective attacks where the relevant space is $F \otimes F$ in a CV protocol.

A QKD protocol is described as a map $\mathcal{E}$:
\begin{align}
\rho_{AB}\mapsto(\mathcal{S}_A,\mathcal{S}_B,\mathcal{C}),
\end{align}
where $\rho_{AB}\in(\mathcal{H}_A\otimes\mathcal{H}_B)^{\otimes n}$ is the $n$-mode bipartite state shared by Alice and Bob at the end of the distribution phase.
$\mathcal{S}_A$ and $\mathcal{S}_B$ are Alice and Bob's final keys, and $\mathcal{C}$ corresponds to a transcript of all classical communication as well as Alice and Bob's raw data.

For the security analysis of the DM-CVQKD protocol, we follow the definition of Ref.\cite{mullerquade2009composability}.
A list of properties we expect an \textit{ideal} protocol to have is first given, then the security of \emph{real}
protocol is defined by indistinguishability from the ideal case. Normally, these properties are defined as \textit{secrecy}, \textit{correctness} and \textit{robustness}. The protocol will be considered
as \textit{robust} if it outputs nontrivial keys when the adversary is passive. It means the adversary does disturb quantum communications. We consider here the strongest type of security by exposing the protocol to \emph{general attacks}. This means that an adversary may arbitrarily tamer with the signals exchanged between Alice and Bob over quantum channels\cite{cai2009finite-key}. Moreover, classical channels may be
monitored by the adversary. In this paper, we first prove the security of the four-state protocol against \textit{Gaussian collective attacks} and then reduce the security proof of the \textit{general attacks} to that of \textit{collective attacks} with Gaussian \textit{de Finetti reduction} \cite{PhysRevLett.118.200501}.

A protocol is \textit{correct}, if the outputs of the protocol on Alice and Bob's side are identical. It is $\varepsilon_\mathrm{cor}$-correct if it is $\varepsilon_\mathrm{cor}$-indistinguishable from a correct protocol, from which we have $\mathrm{Pr}[S_A \ne S_B] \leq \epsilon_{\mathrm{cor}}$\cite{tomamichel2012tight}. A protocol is \textit{secrecy} if the protocol generates a key $S$ that is uniformly distributed and
independent of the states of the system $E$ held by the adversary. A key is $\epsilon$-secret from $E$ if it is $\epsilon$-close to a uniformly distributed key that is uncorrelated with the
adversary \cite{PhysRevLett.98.140502}, that is
\begin{align}
\min_{\rho_E}{\frac{1}{2}\|\rho_{SE} -\omega_{S}\otimes\rho_{E}\|_1}\leq\epsilon,
\end{align}
where $\omega_{S}$ is the completely mixed states on $S$. Correspondingly, a QKD protocol is secret, if, for any attack strategy, $\epsilon=0$ whenever the protocol outputs a key.
It is $\epsilon_\mathrm{sec}$-secret, if its output key is $\epsilon$-secret with $(1-p_\mathrm{abort})\epsilon\leq\epsilon_\mathrm{sec}$, where $p_\mathrm{abort}$ is the probability that the protocol aborts.
This probability depends on the strategy of the adversary, that is on the input state $\Psi_{ABE}$. Indeed, the adversary can always choose to cut the line between Alice and Bob and hence always make the protocol abortifacient. It is fine since the key will nevertheless be secure. An important parameter is the \textit{robustness}, $\epsilon_{\mathrm{rob}}$, of the protocol, which corresponds to the abortion probability if the adversary is passive and if the characteristics of quantum channels are conformed to what was expected. For instance, in the case of a DM-CVQKD protocol, a typical quantum channel is characterized by its transmittance $T$ and excess noise $\xi$.
A QKD protocol is $\epsilon$-secure if it is $\epsilon_{\mathrm{sec}}$-secret and $\epsilon_{\mathrm{cor}}$-correct with $\epsilon_{\mathrm{sec}} + \epsilon_{\mathrm{cor}} \leq \epsilon$.

An operational way of quantifying the security is to bound the diamond distance between two CPTP maps $\mathcal{E}$ and $\mathcal{F}$, where $\mathcal{E}$ denotes the practical DM-CVQKD protocol and $\mathcal{F}$ is an ideal version of the protocol. The ideal version can be obtained by concatenating $\mathcal{E}$ and a (virtual) protocol $\mathcal{P}$ that replaces the final keys $S_A$ and $S_B$ by a perfect key $S$.
We denote $\mathcal{F}=\mathcal{P}\circ\mathcal{E}$ and the protocol $\mathcal{E}$ is $\epsilon$-secure if $\frac{1}{2}\|\mathcal{E}-\mathcal{F}\|_\diamond\leq\epsilon$, which will be detailedly analyzed in Sec.\ref{de finetti reduction}.

In the analysis of the finite regime, symmetrization is a crucial property when analyzing the diamond distance. The diamond distance can be bounded by computing the distance when applied to an
independent and identically distributed (i.i.d) state when the symmetrization property is satisfied. This is the idea behind the Postselection technique\cite{PhysRevLett.102.020504},
which shows that for such protocols, collective attacks are asymptotically optimal. The protocol is invariant under permutations of the $N$ particle pairs held by Alice and Bob after the distribution phase.
Besides, the rotation symmetrization as well as active symmetrization is also considered in the next section. Normally, the operation is applied on classical
data which corresponds to the quantum symmetrization since symmetrization is a costly process that one would like to avoid in a practical implementation.
We also assume that the measurement devices of Alice and Bob are trusted and behave accordingly to their theoretical model. In the four-state protocol, the state sent to Bob in the prepare-and-measure scheme is a mixture of four coherent states $\rho=\frac{1}{4}\sum^3_{k=0}|\alpha_k\rangle\langle\alpha_k|$ with $\alpha_k=\alpha e^{i(2k+1)\frac{\pi}{4}}$. To avoid too much complication, we also assume that Alice's state preparation and Bob's detection are ideal.

\section{Description of the DM-CVQKD protocol}

In this section, we devote to the security analysis of the DM-CVQKD protocol, denoted by $\mathcal{E}_0$, for which we prove composable security against collective attacks. We focus on the EB version of the four-state protocol as the PM protocol is usually used for guiding the experiment while the EB protocol is used for security analysis. In order to prove security in terms of general attacks, one needs to add another step to the protocol, involving an energy test which will be illustrated in Sec.\ref{de finetti reduction}.

\textbf{Security analysis.} We now present one main result of this work. It states that the DM-CVQKD protocol mentioned above is both $\epsilon_\mathrm{cor}$-correct and $\epsilon_\mathrm{sec}$-secret if the length $l$ of the secret key $S_A$ is appropriately selected for a given set of observed value. See \textbf{Box 1} for the different parameters.

If the length $l$ of bit string $S$, which forms the secret key $S_A$, satisfies the constraint
\begin{align}
 l\leq 2n[2\hat{H}_{\mathrm{MLE}}(U)-f(\Sigma^{\mathrm{max}}_a,\Sigma^{\mathrm{max}}_b,\Sigma^{\mathrm{min}}_c)]-\mathrm{leak}_\mathrm{EC}-\Delta_{\mathrm{AEP}}-\Delta_{\mathrm{ent}},
\end{align}
the protocol is $\epsilon$ secret with $\epsilon=\epsilon_\mathrm{PE}+\epsilon_\mathrm{sm}+\epsilon_\mathrm{ent}+\epsilon_\mathrm{cor}$, where $\hat{H}_\mathrm{MLE}(U)$ denotes the empiric entropy of $U$, $\mathrm{leack}_\mathrm{EC}=\lceil\log_2{\frac{1}{\epsilon_\mathrm{cor}}}\rceil$, $\Delta_{\mathrm{AEP}}:=\sqrt{n}(16+\log\frac{2}{\epsilon_\mathrm{sm}^2}+8\sqrt{\log\frac{2}{\epsilon_\mathrm{sm}^2}})+4\frac{\epsilon_\mathrm{sm}}{p}+\log\frac{2}{p^2}$, $\Delta_{\mathrm{ent}}:=n\log n\sqrt{2\log\frac{2}{\epsilon_\mathrm{ent}}}$
and $f$ is the function computing the Holevo information between Eve and Bob's measurement results of a Gaussian state with covariance matrix parametrized by $\{\sum^{\mathrm{max}}_a,\sum^{\mathrm{max}}_b,\sum^{\mathrm{min}}_c\}$.
The function $f$ is defined as:
\begin{align}
f(x,y,z):=g(\nu_1)+g(\nu_2)-g(\nu_3),
\end{align}
where $\nu_1$ and $\nu_2$ are the symplectic eigenvalue of the covariance matrix
$\left[
\begin{smallmatrix}
x I_2 & z \sigma_z  \\
z \sigma_z  &  y I_2 \\
\end{smallmatrix}
\right]$,
$\nu_3=x-\frac{z^2}{1+y}$, $\sigma_z=\mathrm{diag}(1,-1)$ and the entropy function $g(x):=\frac{x+1}{2}\log_2{\frac{x+1}{2}}-\frac{x-1}{2}\log_2{\frac{x-1}{2}}$.

\begin{figure}[htbp]
\textbf{Box 1}: The protocol $\mathcal{E}_0$ which mainly follow the discussion in \cite{PhysRevA.83.042312}.
\begin{framed}
  \centering
\begin{enumerate}
\item \textbf{State Preparation:} Alice draws a string four-dimensional random vectors, each chosen from one of the three following distribution: random vectors on the three-dimensional
    sphere with the appreciate radius, random vector on the three-dimensional sphere with an appropriately fluctuating radius, or Gaussian vectors used for parameter estimation.
\item \textbf{Symmetrization:} Alice randomly draws a random orthogonal transformation $R$ from the orthogonal group $O(4n)$. $R$ can be chosen uniformly in a well-chosen subset
of $O(2n)$ which has the advantage of allowing for efficient descriptions of its elements. Alice computes the vector $x'=Rx$, which is the image of $x$ by the orthogonal transformation
$R$, and use this vector for modulation.
\item \textbf{Distribution:} Alice sends these states over quantum channel to Bob. The channel is assumed to be linear and is modeled by a transmission $T$ and a excess noise $\xi$.
\item \textbf{Measurement:} Bob receives the states after the quantum channel and measures them, with a heterodyne detection. He obtains a $4n$-dimensional vector $y'$.
    Alice describes $R$ to Bob through the classical authenticated channel. Bob applies $R^{-1}$ to his vector $y'$ and obtains $y=R^{-1}y'$.
\item \textbf{Data Filtering:} Alice reveals which subsets should be kept for the key distillation, which ones should be discarded (decoy states) and which ones should be used for parameter estimation (Gaussian states).
\item \textbf{Error Correction:} Bob sends some side information to Alice over the classical channel, typically the syndrome $U$ of his string relative to a binary code they agreed on beforehand. Alice outputs a guess $U_A$ for the string of Bob. Bob computes a hash of $U$ of length $\lceil \log(1/\epsilon_{\mathrm{cor}}) \rceil$ and sends it to Alice who compares it with her own hash. If both hashes coincide, the protocol resumes, otherwise it aborts. The value $\mathrm{leak}_{\mathrm{EC}}$ corresponds to the total number of bits sent by Bob during the error correction phase.
\item \textbf{Parameter Estimation:} Bob sends $n_{\mathrm{PE}}$ bits of information to Alice that allow her to obtain three values $\gamma_a, \gamma_b$ and $\gamma_c$ defined in the following equation.
    If $\gamma_a \leq \Sigma_a^{\max}$, $\gamma_b \leq \Sigma_b^{\max}$ and $\gamma_c \geq \Sigma_c^{\min}$, then the protocol continues. Otherwise it aborts.
\item \textbf{Privacy Amplification:} Alice and Bob apply a random universal$_2$ hash function to their respective strings, obtaining two strings $S_A$ and $S_B$ of size $l$.
\end{enumerate}
\end{framed}
\label{box1}
\end{figure}

In order to calculate the symplectic eigenvalues $\nu_{1,2}$ of the two-mode covariance matrix
\begin{align}
\gamma_{AB}=\left[
\begin{matrix}
x I_2 & z \sigma_z  \\
z \sigma_z  &  y I_2 \\
\end{matrix}
\right],
\end{align}
we need to define a second symplectic invariant,
\begin{align}
\Delta:={\nu_1}^2+{\nu_2}^2=\det xI_2+\det yI_2+2\det z\sigma_z,
\end{align}
where $
\det\gamma_{AB}:={\nu_1}^2{\nu_2}^2$. The symplectic eigenvalues are solutions of the second order polynomial
\begin{align}
z^2-\Delta z+\det\gamma_{AB}=0,
\end{align}
which gives the solution $\nu^2_{1,2}=\frac{1}{2}[\Delta\pm\sqrt{\Delta^2-4\det\gamma_{AB}}]$.

\subsection{State Preparation}

In the DM protocol, Gaussian modulation is used for parameter estimation procedure. Unfortunately, it is not a priori possible to use two different modulations for key distribution and
parameter estimation. We add a $third$ modulation consisting of decoy states to defend the eavesdropper. We call $``key"$, $``decoy"$, and $``G"$ the modulations correspondingly, respectively,
to states used for the key distillation, decoy states, and states used for parameter estimation purpose. The three states are defined as \cite{PhysRevA.83.042312}:
\begin{align}
\sigma^d_{key}&=\int p_{key}(\alpha)|\alpha\rangle\langle\alpha|d\alpha,\\
\sigma^d_{decoy}&=\int p_{decoy}(\alpha)|\alpha\rangle\langle\alpha|d\alpha,\\
\sigma^d_G&=\int p_G(\alpha)|\alpha\rangle\langle\alpha|d\alpha,
\end{align}
where $\alpha\in\mathbb{R}^d$ for the $d$-dimensional protocol. The probability distribution $p_{decoy}$, to make the states used for parameter estimation indistinguishable from that used to distill
a key, is chosen such that
\begin{align}
p\sigma^d_{key}+(1-p)\sigma^d_{decoy}=\sigma^d_G,
\end{align}
with $p=p^\mathrm{succ}_d$. After the exchange of quantum states is complete, Alice announces to Bob which states can be used for the key, which states can be discarded and which states
can be used for parameter estimation. We mainly discuss the states used for key distillation.

Alice prepares the four coherent states $|\alpha e^{i(2k+1)\pi/4}\rangle$ with $k\in\{0,1,2,3\}$. The amplitude $\alpha$ (taken as a real number) is chosen so as to maximize the secret key rate
according to the expected experimental parameters involving the transmission of line and excess noise. The prepared states can be expressed as a mixture of the four coherent states $\rho=\frac{1}{4}\Sigma^3_{k=0}|\alpha_k\rangle\langle\alpha_k|$, where $\alpha_k=\alpha\exp(i(2k+1)\pi/4)$. The EB version makes use of a purification $|\Phi\rangle$ of this state given by
$\rho=\textmd{tr}_A(|\Phi\rangle\langle\Phi|)$, which can be diagonalized as
\begin{align}\label{eq13}
\rho=\lambda_0|\phi_0\rangle\langle\phi_0|+\lambda_1|\phi_1\rangle\langle\phi_1|+\lambda_2|\phi_2\rangle\langle\phi_2|+\lambda_3|\phi_3\rangle\langle\phi_3|,
\end{align}
where $\lambda_{0,2}=\frac{1}{2}e^{-\alpha^2}[\cosh(\alpha^2)\pm\cos(\alpha^2)]$, $\lambda_{1,3}=\frac{1}{2}e^{-\alpha^2}[\sinh(\alpha^2)\pm\sin(\alpha^2)]$
and
$
|\phi_k\rangle=\frac{e^{-\alpha^2/2}}{\sqrt{\lambda_k}}\sum^\infty_{n=0}\frac{\alpha_{4n+k}}{\sqrt{(4n+k)!}}(-1)^n|4n+k\rangle
$
for $k\in\{0,1,2,3\}$. A particular purification of $\rho$ is
\begin{align}
|\Phi\rangle=\sum^3_{k=0}\sqrt{\lambda_k}|\phi_k\rangle|\phi_k\rangle=\frac{1}{2}\sum^3_{k=0}|\psi_k\rangle|\alpha_k\rangle,
\end{align}
where
$
|\psi_k\rangle=\frac{1}{2}\sum^3_{m=0}e^{-i(1+2k)m(\frac{\pi}{4})}|\phi_m\rangle
$
are orthogonal discrete-modulation states.

The EB version of the four-state protocol is described as follows. Alice prepares the entangled state and performs the projective measurement
$\{|\psi_0\rangle\langle\psi_0|,|\psi_1\rangle\langle\psi_1|,|\psi_2\rangle\langle\psi_2|,|\psi_3\rangle\langle\psi_3|\}$ on her half, thus preparing the coherent state $|\alpha_k\rangle$ with the measurement result corresponding to the $k_{th}$ state.
The covariance of DM-CVQKD is given by
\begin{align}
\gamma=\left[
\begin{matrix}
(V_A+1)\mathbb{I}_2 & Z\sigma_z  \\
Z\sigma_z  &  (V_A+1)\mathbb{I}_2 \\
\end{matrix}
\right],
\end{align}
 where $V_A=2\alpha^2$ is the variance of the signal while $\mathbb{I}$ and $\sigma_z$ is defined as before.
For the four-state protocol, one has the parameter
\begin{align}\label{eqZ}
Z=V_A(\frac{\lambda^{3/2}_0}{\lambda^{1/2}_1}+\frac{\lambda^{3/2}_1}{\lambda^{1/2}_2}+\frac{\lambda^{3/2}_2}{\lambda^{1/2}_3}+\frac{\lambda^{3/2}_3}{\lambda^{1/2}_0}).
\end{align}

\subsection{Distribution}
Alice sends the prepared states through a quantum channel to Bob who measures either one of quadratures with a homodyne or both quadratures with heterodyne detection.
In this paper, we consider a situation that the quantum channel is exposed to the eavesdropper, Eve. The quantum state is described by an arbitrary
density operator $\hat{\rho}^{\otimes n}\in\mathcal{D}(\mathcal{H}^{\otimes n})$. The N-mode quantum channel is a linear map $\mathcal{E}:
\hat{\rho}^{\otimes n}\rightarrow\mathcal{E}(\hat{\rho}^{\otimes n})\in\mathcal{D}(\mathcal{H}^{\otimes n})$.
A multimode quantum channel can be represented by a unitary interaction $U$ between an input state $\hat{\rho}$ and a pure state $|\Phi\rangle_E$ of ancillary $N_E$ modes
coupling with the environment. After that, the output of the channel can be achieved by tracing out the environment after interaction.
We note that in the physical representation provided by Stinespring dilation, there is also an output of the environment. Here, we consider the complementary quantum channel $\hat{\mathcal{E}}: \hat{\rho}^{\otimes n}\rightarrow\hat{\mathcal{E}}(\hat{\rho}^{\otimes n})$ which is obtained by tracing out the system after interaction.

We assume that the quantum channel is linear and its input-output relations of the quadrature operators in Heisenberg representation are given by
\begin{align}
X_\mathrm{out}=g_XX_\mathrm{in}+B_X, \quad P_\mathrm{out}=g_PP_\mathrm{in}+B_P,
\end{align}
where $B_X$ and $B_P$ denote the additional noises of the environment which are uncorrelated with the input quadratures $X_\mathrm{in}$ and $P_\mathrm{in}$.
Such relations have been extensively used, for instance, in the context of quantum non-demolition measurements of continuous variables, which are related to the linearized approximation
commonly used in quantum optics. Gaussian channels (whose inputs and outputs are all Gaussian state) are an example of the linear quantum channels.
In particular, some linear quantum channels may be non-Gaussian when the additional noises $B_X$ and $B_P$ don't obey Gaussian distribution.

In a linear quantum channel, the transmission coefficients are characterized by $T_X=g^2_X$ and $T_P=g^2_P$, and the variance of the additional noises are characterized by $B_X$ and $B_P$.
These quantities remain held when the modulation of Alice is discrete modulation, the the measured value is the same as that of the Gaussian modulation (as these values are intrinsic properties of the channel).
The covariance matrix can be determined with the known information and Eve's information can be bounded with the Gaussian optimality theorem.
The covariance matrix $\Gamma_{AB}$ of the state in the entanglement-based is given by
\begin{align}
\Gamma_{AB}=\left[
\begin{matrix}
(V_A+1)\mathbb{I}_2 & \sqrt{T}Z\sigma_z  \\
\sqrt{T}Z\sigma_z  &  (TV_A+1+T\xi)\mathbb{I}_2 \\
\end{matrix}
\right],
\end{align}
where $V_A$, $T$ and $\xi$ represent Alice's variance, the channel transmission and the excess noise evaluated experimentally in the prepare and measure scenario.

\subsection{Measurement}

In the EB DM-CVQKD protocol, Alice and Bob each has access to $2n$ modes.
The $2n$ coherent states kept by Alice can be expressed as $\{|X^r_A+iP^r_A\rangle\}_{\{r=1,2, \cdots, 2n\}}$ and those sent to Bob are $\{|X^r_B+iP^r_B\rangle\}_{\{r=1,2, \cdots, 2n\}}$, respectively.
These modes are detected by heterodyne detection and the $4n$ modes are achieved,
corresponding to the $2n$ position quadratures $\{X^{1}_k, X^{2}_k, \cdots, X^{2n}_k\}$ and the $2n$ phase quadratures $\{P^{1}_k, P^{2}_k, \cdots, P^{2n}_k\}$,  $\forall k\in\{A,B\}$.
The relationships between Alice and Bob's quadratures can be given by
\begin{align}
X^r_A=\frac{V_A}{(V_A+1)^2}\sum^3_{k=0}\frac{\lambda^{3/2}_{k-1}}{\lambda^{1/2}_{k}}\cdot X^r_B,  \quad P^r_A=-\frac{V_A}{(V_A+1)^2}\sum^3_{k=0}\frac{\lambda^{3/2}_{k-1}}{\lambda^{1/2}_{k}}\cdot P^r_B,
\end{align}
 where $V_A$ is the variance of the coherent state and $\lambda_k$ can be calculated from Eq.(\ref{eq13}).

In the equivalent prepare-and-measure version of the DM-CVQKD protocol, Alice doesn't need to possess and measure coherent states.
In experiments, Alice prepares the $2n$ coherent states $\{|X^r_B+iP^r_B\rangle\}_{\{r=1,2, \cdots, 2n\}}$ and sends them to Bob.
Alice keeps data $\{X^1_A, X^2_A, \cdots, X^{2n}_A\}$ and $\{P^1_A, P^2_A, \cdots, P^{2n}_A\}$ in her laboratory rather than coherent states $\{|X^1_A+iP^1_A\rangle, |X^2_A+iP^2_A\rangle, \cdots, |X^{2n}_A+iP^{2n}_A\rangle\}$,
where the amplitude $\alpha$ is chosen so as to maximize the secret key rate with the expected experimental parameters (transmittance $T$ and excess noise $\xi$).

After receiving Alice's signals, Bob performs heterodyne detection on them and obtains two string real random variables $\{X^r_B\}$ and $\{P^r_B\}$, $\forall r\in\{1,2,\cdots, 2n\}$.
Alice and Bob apply a reverse reconciliation to deal with these codes. The signs $b_i$ of $X_B$ and $P_B$ are used for encoding the raw key bit. Namely, we have
\begin{align}
b_i := \left\{
\begin{array}{cccc}
11  &  \!\textrm{if}&   \!\!\textrm{$X_B>0$ and $P_B$>0,} \\
01  &  \textrm{if } &\textrm{$X_B<0$ and $P_B>0$,} \\
00  &  \textrm{if }& \textrm{$X_B<0$ and $P_B<0$,} \\
10  &  \textrm{if }& \textrm{$X_B>0$ and $P_B<0$.}
\end{array}
\right.
\end{align}
All Alice needs to do is to recover the string $\textbf{b}=(b_1, b_2, \cdots, b_{2n})$. In order to help Alice correct her data, Bob sends some side information over classical channels, typically the syndrome of his string
relative to a binary code they agreed on beforehand.

\subsection{Active Symmetrization}
A protocol is said to be invariant under some set of transformation $\mathcal{G}$ if for any element $g\in\mathcal{G}$, there exits a completely positive trace-preserving (CPTP) map $\mathcal{K}_g$ such that
\begin{align}
\mathcal{E}\circ g=\mathcal{K}_g\circ\mathcal{E}.
\end{align}
If we consider here for $\mathcal{G}$ the group of conjugate passive symplectic operations applied on Alice's $n$ modes and Bob's $n$ modes, then for a given operation $g$ applied to the states,
the map $\mathcal{K}_g$ is obtained by applying the orthogonal transformations corresponding to $g$ on the classical data measured by Alice and Bob. If the protocol is invariant under the whole
group, then it is sufficient to look at Gaussian states to prove security against collective attacks \cite{PhysRevA.81.062314}. One simple way to ensure that a protocol is invariant under a set
$\mathcal{G}$ of transformations is for Alice and Bob to actively apply random transformations of $\mathcal{G}$ to their states. Particularly, the security can be done by assuming they are sharing
Gaussian states in the entangled version of the protocol if Alice and Bob both apply random orthogonal transformations to their classical vectors in the prepare-and-measure scheme.
The goal of symmetrization in security analysis is to make sure that the state shared by Alice and Bob is as isotropic as possible. In addition, the symmetrization also plays an important role
in preventing Eve's attack.

We apply an active symmetrization step to the state $\rho_{AB}$ before the protocol to make sure that the protocol is invariant under specific transformations \cite{PhysRevA.83.042312}.
To perform the active symmetrization, we choose a subset $\mathcal{F}$ of the orthogonal group for Alice and Bob, and to apply the same element $f\in\mathcal{F}$ to their data before
starting the postprocessing. We take $\mathcal{F}$ to be a subset of the orthogonal group with following properties as taking for $\mathcal{F}$ the whole orthogonal group is not necessary:
drawing a random element $f$ from the uniform measure on $\mathcal{F}$ should be doable with resources scaling at most linearly in $n$, the description of $f$ should also be linear in $n$,
and applying $f$ (or $f^{-1}$) to a random vector of $\mathbb{R}^n$ should also be at most linear in $n$ \cite{PhysRevA.83.042312}.
Moreover, the symmetrization should work as well as possible, meaning that $\mathcal{F}$ should symmetrize the state as much as possible. Construction of such possible subsets $\mathcal{F}$ can be
found in \cite{PhysRevA.83.042312}.

\subsection{Error Correction}

From the classical communication perspective, error correction is referred to channel coding for the so-called additive white Gaussian noise (AWGN) channel, where a binary modulation is usually sent over
an AWGN channel. The present protocol can thus be seemed as a hybrid between the Gaussian modulation protocol, with which it shares the physical implementation
as well as the security proofs based on the optimality of Gaussian states, and the DM protocol  with post-selection, for which error correction is
substantially convenient to perform. The error correction procedure aims to help Alice learn the string $U$. A possible technique to perform error correction is called the low-density
parity-check (LDPC) codes. However, the LDPC codes are not universal in a sense that they have not been optimized for each channel. A special LDPC code, the multi-edge type LDPC code, was developed for the low signal-to-noise ratio (SNR). Such a code displays good performance with low rates even though it could not solve our problem completely.
The practical transmission rate $R$ is linked to the reconciliation efficiency $\beta$ with
\begin{align}
\beta=\frac{R}{C_\text{Gauss}},
\end{align}
where
$
C_\text{Gauss}=\frac{1}{2}\log_2(1+s)
$
is the capacity of the AWGN channel (which is achieved with a Gaussian modulation) and $s$ is the SNR. In this protocol, this capacity can not be reached as we consider the binary modulation.
Similarly, the maximal value of mutual information between Alice and Bob is given by the capacity of the AWGN channel:
\begin{align}
C_\text{BI-AWGN}(s)=-\int\phi_s(x)\log_2(\phi_s(x))dx-\frac{1}{2}\log_2(2\pi e)+\frac{1}{2}\log_2(s),
\end{align}
where
\begin{align}
\phi_s(x)=\sqrt{\frac{s}{8\pi}}(e^{s(x+1)^2/2}+e^{-s(x-1)^2/2}).
\end{align}
Moreover, we rewrite the reconciliation efficiency of the DM protocol which is used for assessing the quality of the error correction as
\begin{align}
\beta_\text{discrete}:=\beta_\text{modulation}\frac{R}{C_\text{BI-AWGN}},
\end{align}
 with
$
\beta_\text{modulation}=\frac{C_\text{Gauss}}{C_\text{BI-AWGN}}.
$
Therefore, the reconciliation efficiency can be calculated as
\begin{align}
\beta_\text{discrete}=\frac{(4n-{\textmd{leak}_{EC}})2n\log_2(1+s)}{C^2_\text{BI-AWGN}},
\end{align}
where the SNR is that of the expected linear quantum channel mapping $X$ to $Y$. The reconciliation shows how much information was extracted through the error correction procedure,
comparing to the available mutual information corresponding to a linear quantum channel of transmittance $T$ and excess noise $\xi$. Particularly, we have:
\begin{align}
s=\frac{TV_A}{2+T\xi}.
\end{align}
The quantity $\frac{1}{2}\log_2(1+s)$ is the mutual information for each use of the channel, leading to an upper bound on the quantity of classical information that
can be transmitted over the classical channel $X_i\mapsto Y_i$. The reconciliation efficiency therefore gives the distance between the error correction procedure and the ideal one with $\beta=1$.

We show an approach to generate a code of rate $R/k$ out of a code of rate $R$ at the end of quantum exchange. Alice and Bob share the two correlated vectors
$x=(x_1, x_2, \cdots, x_N)$ with $x_i=\pm\alpha/\sqrt{2}$ and $y=(y_1, y_2, \cdots, y_N)$. We use the concatenation
of a capacity-achieving code $C$ of length $m$ and a repetition code of length $k$. Bob starts by defining the vector $Y=(Y_1, Y_2, \cdots, Y_m)$ where $Y_i=\textrm{sig}(y_{k(i-1)+1})$
for $i\in \{1, 2, \cdots, m\}$ with $N=mk$,. The goal of the reconciliation is for Alice to compute the vector $Y$. To do this, Bob sends some side information, the vector $\{|y_1|,|y_2|,\cdots,|y_N|\}$,
the $m$ vectors $\{(1,\textrm{sgn}{y_{k(i-1)+1}}\times \textrm{sgn}{y_{k(i-1)+2}}, \cdots, \textrm{sgn}(y_{k(i-1)+1}\times y_{ki}))\}$, and the syndrome of $Y$ for the code $C$. This scheme allows
Alice and Bob to extract $m$ bits out of their initial $N=km$ data.
After this step, it is necessary to judge whether $X=Y$ or not. To achieve this, the usual technique is for Alice and Bob to choose a random universal hash function mapping $4n$ bit strings
to string of length $\lceil\log(1/\epsilon_\text{cor})\rceil$. When Bob reveals his hash to Alice, the protocol resumes if both hashes coincide and will be aborted otherwise.
The length of the hash is chosen so that the protocol is $\epsilon_\text{cor}$-correct.

\subsection{Parameter Estimation}

In this section, we suggest the parameter estimation procedure for the DM-CVQKD protocol. Furthermore, taking advantage of specific symmetries of the protocols in phase space \cite{leverrier2009security},
it is possible to reduce this number to 3, the variances of Alice and Bob's reduced states and the variance.
It is a coarse-grained version of quantum tomography because the parameter estimation only relays on a small number of the parameters of the
underlying quantum state. On the positive side, the covariance matrix can be symmetrized through a technique similar to the one explained in \cite{leverrier2009security}.
To implement the parameter estimation, we show that quantum state tomography, together with an appropriate data analysis procedure, yields the reliable and tight error bounds,
specified in terms of confidence regions \cite{christandl2011reliable}. Confidence regions are subsets of the state space in which the real state exists with high probability,
independently of any prior assumption on the distribution of the possible states.

The quantum state tomography is applied by measuring a portion of quantum states in the $k$-dimension Hilbert space, together with an appropriate data analysis procedure, resulting in a reliable and tight error bounds $\epsilon_\mathrm{PE}$ and confidence regions. $\epsilon_\mathrm{PE}$ denotes the probability of discarding the protocol. The protocol will
be abandoned if $\epsilon_\mathrm{PE}=1$. From this point of view, we would like to set $\epsilon_\mathrm{PE}$ as small as possible to make the protocol more robust. On the other hand,
the region $\mathcal{R}$ becomes the total space, $\mathcal{H}$, if $\epsilon_\mathrm{PE}$ is set to be zero. For the protocol, the size of the region will influence the tightness of
the bound of the secret key rate. The large confidence regions may lead to failure of detecting eavesdropper, whereas the smaller confidence regions lead to the tight key rates.
It is necessary to find out a balance of \textit{secrecy} and \textit{robustness}. The probability of abandoning the protocol, $\epsilon_\mathrm{PE}$, is an important parameter that can be given by
\begin{align}
\forall \rho^{n+k} \in\mathcal{H}^{\otimes (n+k)}, \mathrm{Pr}\left[ \rho^n \in R \right] \geq 1-\epsilon_{\mathrm{PE}},
\end{align}
where $\rho^n$ and $R$ are the output of the map tomography applied to $\rho^{n+k}$. $\rho^n$ includes $n$ sets of data and $R$ is a confidence region obtained from $k$ sets of data.

Subsequently, we estimate the covariance matrix of a bipartite state $\tilde{\rho}^n_{AB}$ that is obtained from $\rho^{2n}_{AB}$ after a rotation symmetrization,
characterized by three parameters:
\begin{align}
\Sigma_a & := \frac{1}{2n} \sum_{i=1}^n \left[ \langle X_{A_i}^2 \rangle + \langle P_{A,i}^2\rangle \right]\\
\Sigma_b & := \frac{1}{2n}  \sum_{i=1}^n  \left[\langle X_{B_i}^2 \rangle + \langle P_{B,i}^2\rangle \right]\\
\Sigma_c & := \frac{1}{2n}  \sum_{i=1}^n  \left[\langle X_{A,i} X_{B,i} \rangle - \langle P_{A,i} P_{B,i}\rangle \right]
\label{cov-matrix-elements}
\end{align}
where $X_{A,i}$ and $P_{A,i}$ are the quadrature operators $\frac{1}{\sqrt{2}} (\hat{a}_i + \hat{a}^\dagger_i)$ and $\frac{1}{\sqrt{2}} (\hat{a}_i - \hat{a}^\dagger_i)$ of the $i^{\mathrm{th}}$ mode of Alice.
In the protocol, the assignment of parameter estimation is to find confidence regions of the parameters $\Sigma_a$, $\Sigma_b$, $\Sigma_c$ which are different from the protocol that estimates $t_{\mathrm{min}}$ and $\sigma^2_{\mathrm{max}}$ \cite{PhysRevA.81.062343}. For a given covariance matrix, the state maximizing Eve's information is Gaussian. The only possible loophole would be that there might exist a DM state,
compatible with a Gaussian assumption, which would be attacked by Eve than the Gaussian state estimated by Alice and Bob. This loophole is caused by an inaccurate understanding of covariance matrix in the finite-size scenario
and can be solved by introducing decoy states. Therefore, we conjecture that the Gaussian optimality still holds in a non-asymptotic scenario,
and in the following, we make the assumption of a Gaussian channel. After that, we have the symmetrized matrix that can be used for deriving the secret key rate:
\begin{align}
\Gamma^{\mathrm{sym}} := \bigoplus_{i=1}^n \left[
\begin{matrix}
\Sigma_a & 0 & \Sigma_c & * \\
0 & \Sigma_a & * & -\Sigma_c \\
\Sigma_c & * & \Sigma_b & 0 \\
* & -\Sigma_c & 0 & \Sigma_b \\
\end{matrix}
\right].
\end{align}
The entries will be considered in the procedure of rotation symmetrization.
Based on the properties of Holevo information, it is sufficient to obtain a confidence region of $\{\Sigma_a,\Sigma_b,\Sigma_c\}$ in the forms $[0,\Sigma^\mathrm{max}_a]\times[0,\Sigma^\mathrm{max}_b] \times[\Sigma^\mathrm{max}_c,\infty]$. If the variance of Alice's initial state is $V$, the expected transmittance of the quantum channel is $T$ and the expected excess noise is $\xi$, we have
\begin{align}
&\Sigma^\mathrm{max}_a=V+\delta_a, \\
&\Sigma^\mathrm{max}_b=T(V-1)+1+T\xi+\delta_b, \\
&\Sigma^\mathrm{min}_c=\sqrt{T}(V-1)(\frac{\lambda^{3/2}_0}{\lambda^{1/2}_1}+\frac{\lambda^{3/2}_1}{\lambda^{1/2}_2}+\frac{\lambda^{3/2}_2}{\lambda^{1/2}_3}+\frac{\lambda^{3/2}_3}{\lambda^{1/2}_0})-\delta_c,
\end{align}
where $\delta_a$, $\delta_b$ and $\delta_c$ are chosen to ensure both robustness and high key rate of the system.

The quantum state tomography would be simplified if the initial states satisfy independent and identical distribution of the form $\rho^{n+k}$.
Unluckily, different from the finite-dimensional systems such as DVQKD, whose density matrix elements are bounded, the coefficients of covariance matrix are a priori unbounded.
A solution to this issue is to apply a rotation symmetrization to the state $\rho^{n+k}$ before measuring $k$ subsystems and inferring properties for the remaining $n$ modes.
A new problem appears when the output state of the tomographic procedure will loss the independent and identical distribution structure, which makes the analysis of collective attacks
more complicated. The tool used for solving this problem is to apply quantum de Finetti theorem, from which a class of states in $\mathcal{H}^{\otimes n}$, namely symmetric states,
can be approximated by mixture of independent and identical distributed states.

We consider the security analysis of the DM protocol with collective attacks characterized by the covariance matrix of $\rho_{AB}$. The symmetrization that makes sense of CV QKD is the one
that maximally symmetrizes the state while leaving the parameters of the covariance matrix unchanged. Since we make the assumption of collective attacks, the covariance matrix $\Gamma$
 can be well obtained and estimated by Alice and Bob. The general form of $\Gamma$ is given by
\begin{align}
\Gamma := \left[
\begin{matrix}
\Sigma^{11}_a & \Sigma^{12}_a & \Sigma^{11}_c & \Sigma^{12}_c \\
\Sigma^{21}_a & \Sigma^{22}_a & \Sigma^{21}_c & -\Sigma^{22}_c \\
\Sigma^{11}_c & \Sigma^{12}_c & \Sigma^{11}_b & \Sigma^{12}_b \\
\Sigma^{21}_c & -\Sigma^{22}_c & \Sigma^{21}_b & \Sigma^{22}_b \\
\end{matrix}
\right].
\end{align}
We make use of the rotation symmetrization rather than the traditional permutation symmetrization to transform $\Gamma$ into $\Gamma_{\mathrm{sym}}$.
The reconciliation is always optimized for a Gaussian channel, and hence Bob's data $Y$ can be modeled as $Y=tX+z$, where $t$ is a transmission factor and $z$
is a random variable modeling the added noise with a variance of $\sigma^2$. Therefore, the reconciliation procedure would not be affected
if Alice and Bob perform the same random orthogonal transformation $R\in O(n)$ on their respective data, since one would have $RY=tRX+z'$,
where $z'$ denotes the rotated noise with variance $\sigma^2$ and $R$ takes the form of $\left[
\begin{smallmatrix}
 \sin\theta & \cos\theta \\
-\cos\theta & \sin\theta \\
\end{smallmatrix}
\right]$
with $\theta$ being the rotation angle.
If Alice and Bob apply such a random orthogonal transformation and erase which state has been
performed, they obtain a series of symmetrized data that the covariance matrix takes the form of $\Gamma_{\mathrm{sym}}$. The parameters of matrix $\Gamma$
and $\Gamma_{\mathrm{sym}}$ satisfy the constraints $\Sigma_a=(\Sigma^{11}_a+\Sigma^{22}_a)/2$, $\Sigma_b=(\Sigma^{11}_b+\Sigma^{22}_b)/2$ and $\Sigma_c=(\Sigma^{11}_c-\Sigma^{22}_c)/2$.
Alice and Bob would actually apply the conjugate orthogonal transformations to these position and phase quadratures. The transformation whose $2n\otimes2n$ matrix
in phase space is achieved from the original one by flipping the sign of all rows whose label corresponds to the $p$ quadrature. Therefore, we have
\begin{align}
\begin{array}{cccc}
\mathcal{R}\mathrm{ot}  \mathcal{S}\mathrm{ym}: P_=(P^{\otimes(n+k)}_A\otimes P^{\otimes(n+k)}_B)\rightarrow P_=(P^{\otimes(n+k)}_A\otimes P^{\otimes(n+k)}_B) \\
\rho^{\otimes(n+k)}_{AB} \mapsto [\oplus^{n+k}_{i=1}R_i]\rho^{\otimes(n+k)}_{AB}[\oplus^{n+k}_{i=1}R^T_i]
\end{array}
\end{align}

The above-mentioned symmetrization, which applies orthogonal transformations in phase space rather than permutations in state space, has several characterastics.
It allows the matrix of Alice and Bob to be characterized by the three parameters which can be estimated experimentally.
Besides, it provides enables us an approach of the unconditional security analysis using a de Finetti theorem.

We can derive the confidence interval of the covariance matrix with the yielded quantum states.
For these Gaussian states, $k$ modes are heterodyned for calculating a confidence region of $\Sigma_a$, $\Sigma_b$ and $\Sigma_c$.
Alice and Bob first measure $k$ modes each with heterodyne detection, inferring
a confidence region of the covariance matrix of the $k$ remaining Gaussian states. We take $X_2$ and $Y_2$ denoting the $2k$ vectors of
Alice and Bob's measurement results, respectively. We consider a simple condition that Bob sends his $2k$ measurement results
to Alice through a classical authorized channel for reverse reconciliation. Under the circumstance, it is direct for Alice to compute $\|X_2\|^2$, $\|Y_2\|^2$ and $\langle X_2, Y_2\rangle$
which is useful for the parameter estimation. The procedure of Alice and Bob to exchange their measurement data for
parameter estimation still needs further exploration.

\section{Parameter Estimation}

As explained previous, the aim of parameter estimation is to give confidence regions of $\Sigma_a$, $\Sigma_b$ and $\Sigma_c$ and sift out
these data whose variances pass the test. In a case that $\|X\|^2$, $\|Y\|^2$ and $\langle X,Y\rangle$ are known to Alice and Bob, they can
set about calculating the confidence region.

After Bob receives Alice's states, they pick up these Gaussian states that used for parameter estimation. Alice and Bob split their own $2k$
states into two sets of $k$ modes, denoted as $A_1$, $B_1$ and $A_2$, $B_2$.  $A_1$ and $B_1$ can be used for estimating $A_2$ and $B_2$,  and conversely $A_2$ and $B_2$ can also
Be used for estimate $A_1$ and $B_1$. Without loss of generality, we consider the former situation with
\begin{align}
\|X_1\|^2=\sum_{i=1}^k\alpha^2_{A,i}, \quad
\|Y_1\|^2=\sum_{i=1}^k\alpha^2_{B,i}.
\end{align}
As $\alpha_{A,i}$ and $\alpha_{B,i}$ are with Gaussian distribution, $\frac{\|X_1\|^2}{\alpha^2}$ and $\frac{\|Y_1\|^2}{\alpha^2}$ obey the chi square distribution $(\chi^2(k))$.
In the following, we introduce the lemma of tail bounds of $\chi^2(k)$ distribution.

\begin{lemma}
\label{laurent-massart}
Let $U$ be a $\chi^2$ statistics with $k$ degrees of freedom. For any $x>0$, we have
\begin{align}
\mathrm{Pr}\left[U -k \geq 2\sqrt{kx} + 2x \right] \leq e^{-x}, \quad
\mathrm{Pr}\left[k -U \geq 2\sqrt{kx}  \right] \leq e^{-x}.
\end{align}
\end{lemma}

We consider a case that Alice distributes $\rho_1$ to $\{A_1,B_1\}$ and $\rho_2$ to $\{A_2,B_2\}$, respectively. It is feasible for Alice to compute the confidence region
of $\|X_i\|^2$, $\|Y_i\|^2$ and $\langle X_i,Y_i\rangle$ $(i=1,2)$ as she has knowledge of $\|X\|^2$, $\|Y\|^2$ and $\langle X,Y\rangle$. Therefore, we have
$X=(X_1,X_2)$, $Y=(Y_1,Y_2)$, $\|X\|^2=\|X_1\|^2+\|X_2\|^2$ and $\|Y\|^2=\|Y_1\|^2+\|Y_2\|^2$.
The following is the lemma of building the confidence region of $\|X_1\|^2$, $\|Y_1\|^2$ and $\langle X_1,Y_1\rangle$ based on $\|X\|^2$, $\|Y\|^2$ and $\langle X,Y\rangle$.

\begin{lemma}
\label{estX}
For a given vector $X \in \mathbbm{C}^{2k}$, we have
\begin{align}
&\mathrm{Pr}\left[\|X_1\|^2 \geq \frac{1}{2}\left[1+\frac{5}{2}\sqrt{\frac{\ln (2/\epsilon)}{k}}\right] \|X\|^2 \right] \leq\epsilon \\
&\mathrm{Pr}\left[\|X_1\|^2 \leq \frac{1}{2}\left[1-\frac{11}{5}\sqrt{\frac{\ln (2/\epsilon)}{k}}\right] \|X\|^2 \right] \leq \epsilon,
\end{align}
where $X_1$ is the projection of $X$ on a random subspace of dimension $k$ and $\epsilon \geq 2e^{-k/2}$.
\end{lemma}

\begin{proof}
Attributing to $\|X\|^2=\|X_1\|^2+\|X_2\|^2$, the two variables $\frac{\|X_1\|^2}{\alpha^2}$ and $\frac{\|X_2\|^2}{\alpha^2}$ are subject to the chi square distribution, and hence $\frac{\|X\|^2}{\alpha^2}$
obeys the $\chi^2(2k)$ distribution.
Combining with Lemma \ref{laurent-massart}, we obtain
\begin{align}
\mathrm{Pr}\left[ \frac{\|X_i\|^2}{\alpha^2} \geq k+ 2\sqrt{kx} + 2x \right] \leq e^{-x}, \quad
\mathrm{Pr}\left[ \frac{\|X_i\|^2}{\alpha^2} \leq k- 2\sqrt{kx}  \right] \leq e^{-x}.
\end{align}
Consequently, we have
\begin{align}\label{eqlm2}
\mathrm{Pr}\left[ \frac{\|X_1\|^2}{\|X_1\|^2+\|X_2\|^2} \geq \frac{k+ 2\sqrt{kx} + 2x}{2(k+x)} \right] \leq 2e^{-x}, \quad
\mathrm{Pr}\left[ \frac{\|X_1\|^2}{\|X_1\|^2+\|X_2\|^2} \leq \frac{k- 2\sqrt{kx} }{2(k+x)} \right] \leq 2e^{-x}
\end{align}
After iterating $\gamma=\sqrt{x/k}$ into the right of Eq. (\ref{eqlm2}) for $\gamma\in[0,1]$, we have
\begin{align}
\frac{1+ 2\gamma + 2\gamma^2}{2(1+\gamma^2)} \leq \frac{1}{2}(1+\frac{5\gamma}{2}) \quad \mathrm{and} \quad
\frac{1- 2\gamma}{2(1+\gamma^2)}  \geq \frac{1}{2}(1-\frac{11}{5}\gamma).
\end{align}
\end{proof}

With the relationship $\langle X,Y\rangle=\langle X_1,Y_1\rangle+\langle X_2,Y_2\rangle$ in mind,
we need give the confidence region of $\langle X_i,Y_i\rangle$ due to the data on $\langle X,Y\rangle$.

\begin{lemma}
\label{estXY}
Given two vectors $X, Y \in \mathbbm{R}^{4k}$,  we have
\begin{align}
\mathrm{Pr}\left[\langle X,Y\rangle - (1+\frac{27}{10}\sqrt{\frac{x}{k}}) \left( \|X\|^2 +\|Y\|^2 \right) \leq 4\langle X_i, Y_i\rangle \leq \langle X,Y\rangle + (1+\frac{27}{10}\sqrt{\frac{x}{k}}) \left( \|X\|^2 +\|Y\|^2 \right) \right]  \geq 1- 8 e^{-x},
\end{align}
where $X_1$ and $Y_1$ are the projections of $X$ and $Y$ on a random subspace of dimension $2k$ for any $x \leq k/2$.
\end{lemma}

\begin{proof}
We use Lemma \ref{estX} to prove this lemma with
$
\|X+Y\|^2 = \|X\|^2+2\langle X, Y\rangle+\|Y\|^2
$,
which is the property of 2-norm.
Based on the union bound, except with probability no more than $8e^{-x}$ for $x \leq k$, one has
\begin{align}
\frac{1}{2}\left[1-\frac{11}{5}\sqrt{\frac{x}{k}} \right] \|X+Y\|^2 &\leq  \|X_1 + Y_1 \|^2 \leq \frac{1}{2}\left[1+\frac{5}{2} \sqrt{\frac{x}{k}} \right] \|X+Y\|^2,\\
-\frac{1}{2}\left[1+\frac{5}{2} \sqrt{\frac{x}{k}} \right] \|X-Y\|^2 &\leq  -\|X_1-Y_1 \|^2 \leq -\frac{1}{2}\left[1-\frac{11}{5}\sqrt{\frac{x}{k}} \right] \|X-Y\|^2.
\end{align}
Combing with $\|X\|^2 + \|Y\|^2  \geq 2\langle X, Y\rangle$, we have
\begin{align}
\langle X,Y\rangle - \frac{47}{10}\sqrt{\frac{x}{k}} \left( \|X\|^2 +\|Y\|^2 \right) \leq 2\langle X_i, Y_i\rangle \leq \langle X,Y\rangle + \frac{47}{10}\sqrt{\frac{x}{k}} \left( \|X\|^2 +\|Y\|^2 \right).
\end{align}
\end{proof}

We note in particular that the following bounds hold
\begin{align}
\mathrm{Pr}\left[ \langle X_i, Y_i \rangle \leq \frac{1}{2}\langle X,Y\rangle - \frac{5}{2}\sqrt{\frac{x}{k}}\left( \|X\|^2 +\|Y\|^2 \right) \right] & \leq 4 e^{-x} \label{bound-c},\\
\mathrm{Pr}\left[| \langle X_1, Y_1 \rangle - \langle X_2,Y_2\rangle| \geq 5\sqrt{\frac{x}{k}} ( \|X\|^2 + \|Y\|^2) \right] & \leq 8 e^{-x}.
\end{align}
Subsequently, we have the confidence region of $\|X_2\|$, $\|Y_2\|$ and $\langle X_2, Y_2\rangle$ by using  the confidence region of $\|X_1\|$, $\|Y_1\|$ and $\langle X_1, Y_1\rangle$.

\begin{lemma} For an $8k$-dimensional probability distribution $P(X, Y)$ with $X = (X_1,X_2) \in \mathbbm{R}^{4n}$ and $Y=(Y_1,Y_2) \in \mathbbm{R}^{4k}$, which are symmetrized by rotation operation, we have
\begin{align}
&\mathrm{Pr}\left[ \|X_2\|^2 \geq \left[1+ 4\sqrt{\frac{\log(2/\epsilon)}{2k}}\right] \|X_1\|^2\right] \leq \epsilon, \label{ineqA}\\
&\mathrm{Pr}\left[ \|X_2\|^2 \leq \left[1- 4\sqrt{\frac{\log(2/\epsilon)}{2k}}\right]  \|X_1\|^2\right] \leq \epsilon,\label{ineqB}\\
&\mathrm{Pr}\left[ \langle X_2, Y_2\rangle \leq \langle X_1, Y_1\rangle- 4\sqrt{\frac{\log(2/\epsilon)}{2k}} \left(\|X_1\|^2+\|Y_1\|^2\right) \right] \leq 4\epsilon, \label{ineqC}
\end{align}
\label{X1givesX2} where $\epsilon$ satisfies the constraint ${\frac{\log(2/\epsilon)}{2k}} \leq 0.05$.
\end{lemma}

\begin{proof}
The first two inequalities can be derived from Lemma $\mathrm{II}.1$ in Ref.~\cite{PhysRevLett.110.030502} due to an observation that for $\gamma \in [0, 0.05]$, we have
\begin{align}
\frac{1+2\gamma+2\gamma^2}{1-2\gamma} \geq 1+4\gamma, \quad
\frac{1-2\gamma}{1+2\gamma+2\gamma^2} \leq 1-4\gamma.
\end{align}
Similarly, we have the above inequalities for $\|Y_1\|^2$ and $\|Y_2\|^2$.
Using the strategy as in the proof of Lemma \ref{estXY}, except with probability $8\epsilon$, one can obtain
\begin{align}
\langle X_1,Y_1\rangle-4\sqrt{\frac{\log(2/\epsilon)}{2k}}(\|X_1\|^2+\|Y_1\|^2)\leq\langle X_2,Y_2\rangle\leq\langle X_1,Y_1\rangle+4\sqrt{\frac{\log(2/\epsilon)}{2k}}(\|X_1\|^2+\|Y_1\|^2).
\end{align}
\end{proof}

Now, we show the loophole of the parameter estimation for $[\gamma_a\leq\Sigma^\mathrm{max}_a] \wedge [\gamma_b\leq\Sigma^\mathrm{max}_b] \wedge [\gamma_c\geq\Sigma^\mathrm{min}_c]$.
The problematic cases are the ones where $X_1$ or $X_2$ are not compatible with their own estimation procedure. Followings are the specific cases,
\begin{align}
E_{\mathrm{bad}}^{\|X\|^2} := [\|X_1\|^2 \geq a] \vee [\|X_2\|^2 \geq a] \vee [ (\mathbbm{E} \|X_2\|^2 \geq b) \wedge (\|X_1\|^2 \leq a)]  \vee [ (\mathbbm{E} \|X_1\|^2 \geq b) \wedge( \|X_2\|^2 \leq a)],
\end{align}
where $a$ and $b$ will be optimized later.
Similarly, one has the definition
\begin{align}
E_{\mathrm{bad}}^{\|Y\|^2} := [\|Y_1\|^2 \geq a] \vee [\|Y_2\|^2 \geq a] \vee [ (\mathbbm{E} \|Y_2\|^2 \geq b) \wedge (\|Y_1\|^2 \leq a)]  \vee [ (\mathbbm{E} \|Y_1\|^2 \geq b) \wedge( \|Y_2\|^2 \leq a)]
\end{align}
which is a bad event for the estimation of $B_1$ and $B_2$'s variances.
The bad event for the correlations is given by
\begin{align}
E_{\mathrm{bad}}^{\langle X,Y\rangle} := [\langle X_1, Y_1\rangle \leq c] \vee [\langle X_2, Y_2\rangle \leq c] \vee [ (\mathbbm{E} \langle X_2, Y_2\rangle \leq d) \wedge (\langle X_1, Y_1\rangle \geq c)]  \vee [ (\mathbbm{E} \langle X_1, Y_1\rangle \leq d) \wedge( \langle X_2, Y_2\rangle \geq c)].
\end{align}

\begin{theo}
\label{main-parameter-estimation}
The probability of the bad event $E_{\mathrm{bad}}^{\\|X\|^2} \vee E_{\mathrm{bad}}^{\|Y\|^2} \vee E_{\mathrm{bad}}^{\langle X,Y\rangle}$ is upper bounded by $\epsilon$ for the parameters
\begin{align}
a &  = \frac{1}{2} \left[1+\frac{5}{2}\sqrt{\frac{\log (36/\epsilon)}{k}}\right] \|X\|^2  \quad \text{or} \quad a = \frac{1}{2} \left[1+\frac{5}{2}\sqrt{\frac{\log (36/\epsilon)}{k}}\right] \|Y\|^2,\\
b & =  a \left[ 1 + \frac{432}{\epsilon} \exp[-k/16]\right],\\
c &= \frac{1}{2} \langle X, Y\rangle - \frac{5}{2}\sqrt{\frac{\log(72/\epsilon)}{k}} ( \|X\|^2 + \|Y\|^2),\\
d & = c - 2(\|X\|^2 + \|Y\|^2) \sqrt{\frac{\log(144/\epsilon)}{k}}.
\end{align}
\end{theo}

\begin{proof}
Based on Lemma \ref{X1givesX2}, we define $\epsilon_1(a,y) = 2 \exp\left(-2k\left[ \frac{y-a}{4a}\right]^2 \right)$ such that
$
\mathrm{Pr} \left[ (\|X_1\|^2 \leq a) \wedge ( \|X_2\|^2  \geq y )\right] \leq \mathrm{Pr} \left[ \|X_2\|^2 \geq \frac{y}{a} \|X_1\|^2 \right]
 \leq \epsilon_1(a,y).
$
For $\gamma = b/a$, we  have
\begin{align}
\sum_{\nu=1}^\infty b_{\nu+1} \epsilon_1(a, b_\nu) & = 2a \sum_{\nu=0}^\infty (\gamma  +2 + \nu ) \exp\left[ -\frac{k}{16} (\gamma + \nu)^2  \right]\\
& \leq 2a \exp\left[ -k \gamma^2/16 \right] \sum_{\nu=0}^\infty (\gamma  +2 + \nu ) \exp\left[ -\frac{2k\gamma \nu}{16}  \right]\\
& \leq 8a(\gamma+2) \exp\left[ -k \gamma^2/16 \right],
\end{align}
where the last inequality holds provided that $\exp\left[ -\frac{2k\gamma }{16}  \right] \leq 1/2$.
Using Lemma $8$ in \cite{PhysRevLett.114.070501}, we achieve
\begin{align}
\label{choice-delta}
\mathrm{Pr} \left[ (\|X_1\|^2 \leq a) \wedge ( \mathbbm{E}\|X_2\|^2 \geq a + \delta )\right] \leq \frac{24a}{\delta} \exp\left[ -k/16 \right].
\end{align}
According to Lemma 9 in \cite{PhysRevLett.114.070501},  for $c \geq d$, we obtain
\begin{align}
\mathrm{Pr}\left[ (\mathbbm{E} \langle X_2, Y_2\rangle \leq d - \delta) \wedge (\langle X_1, Y_1\rangle \geq c) \right] &\leq \frac{d}{\delta} \cdot \mathrm{Pr}\left[ (\langle X_2, Y_2\rangle \leq d) \wedge (\langle X_1, Y_1\rangle \geq c) \right]\\
& \leq \frac{d}{\delta} \cdot \mathrm{Pr}\left[ \langle X_1, Y_1\rangle  - \langle X_2, Y_2\rangle\geq c-d \right]\\
& \leq \frac{8d}{\delta}\exp\left[- k \left[\frac{c-d}{5( \|X\|^2 + \|Y\|^2)}\right]^2\right].
\end{align}
Taking $d = c-\delta$, we have
\begin{align}
\mathrm{Pr}\left[ (\mathbbm{E} \langle X_2, Y_2\rangle \leq c -2 \delta) \wedge (\langle X_1, Y_1\rangle \geq c) \right] & \leq \frac{8c}{\delta}\exp\left[- k \left[\frac{\delta}{5( \|X\|^2 + \|Y\|^2)}\right]^2\right].
\label{choice-delta2}
\end{align}
For the suitable values of $a$, $b$, $c$ and $d$ that enables each of the $18$ individual events with probability $\epsilon/18$,
the probability $p_{\mathrm{bad}}^{\mathrm{PE}}$ of the parameter estimation is calculated as
\begin{align}
p_{\mathrm{bad}}^{\mathrm{PE}} \leq \mathrm{Pr} \left[ E_{\mathrm{bad}}^{\|X\|^2}\right] +\mathrm{Pr} \left[ E_{\mathrm{bad}}^{\|Y\|^2}\right]  + \mathrm{Pr} \left[ E_{\mathrm{bad}}^{\langle X,Y\rangle} \right]   \leq \epsilon.
\end{align}
This is achieved for the values of $a$, $b$, $c$ and $d$ given by
\begin{align}
a &  = \frac{1}{2} \left[1+\frac{5}{2}\sqrt{\frac{\log (36/\epsilon)}{k}}\right] \|X\|^2,\\
b & = a \left[ 1 + \frac{432}{\epsilon} \exp[-k/16]\right],\\
c &= \frac{1}{2} \langle X, Y\rangle - \frac{5}{2}\sqrt{\frac{\log(72/\epsilon)}{k}} ( \|X\|^2 + \|Y\|^2),\\
d & = c - 2(\|X\|^2 + \|Y\|^2) \sqrt{\frac{\log(144/\epsilon)}{k}},
\end{align}
where the second equality is derived from Eq.~\ref{choice-delta}, the third equality is from Eq.~\ref{bound-c} and the last equality is from Eq.~\ref{choice-delta2}.
\end{proof}

Finally, we have the parameters
\begin{align}
\gamma_a &:= \frac{1}{2k} \left[ 1 + 3\sqrt{\frac{\log(36/\epsilon_{\mathrm{PE}})}{k}}\right] \|X\|^2-1,\\
\gamma_b &:= \frac{1}{2k} \left[ 1 + 3\sqrt{\frac{\log(36/\epsilon_{\mathrm{PE}})}{k}}\right] \|Y\|^2-1,\\
\gamma_c &:= \frac{1}{2k} \langle X, Y\rangle - 6 \sqrt{\frac{\log (144/\epsilon_{\mathrm{PE}})}{k^3}}(\|X\|^2 + \|Y\|^2),
\end{align}
where a regime of $n$ satisfies the constraint
\begin{align}
 \left[1+\frac{5}{2}\sqrt{\frac{\log (36/\epsilon)}{k}}\right]  \left[1+\frac{360}{\epsilon} \exp[-k/16]\right] \leq  1+3\sqrt{\frac{\log (36/\epsilon)}{k}},
\end{align}
which exists in practical implementations.
Due to the fact that
\begin{align}
\sqrt{9 \log(72/\epsilon)} + \sqrt{4\log(144/\epsilon)}\leq \sqrt{2(9 \log(72/\epsilon) + 4\log(144/\epsilon))} \leq 6 \sqrt{\log(144/\epsilon)},
\end{align} the parameters $\gamma_A$, $\gamma_B$ and $\gamma_C$ give the confidence region of the covariance matrix when the probability of the bad event in parameter estimation is less than $\epsilon$.

\section{Smooth Min-entropy}

The smooth min-entropy is associated to the secrecy of the protocol with leftover hash lemma. As analyzed in \cite{PhysRevLett.118.200501}, the most general attacks can be reduced to collective Gaussian
attacks by applying the Gaussian de Finetti reduction. In order to analysis the security of the DM-CVQKD protocol, we first restrict the attacks to collective attacks. Moreover, the conditional smooth
min-entropy can be estimated using the asymptotic equipartition property (AEP). In this section, we  introduce the leftover hash lemma and use the AEP to calculate the smooth min-entropy.

We consider a random variable $X$ that is partially known to the eavesdropper, Eve, who possesses side information $E$ correlated to $X$. The $Leftover$ $Hash$ $Lemma$ \cite{tomamichel2011leftover} gives the number $l$ of
extractable bits, which is slightly smaller than the min-entropy of $A$ conditioned on $B$, denoted $H_\mathrm{min}(A|B)$.
The relationship between $l$ and $H_{\mathrm{min}}(A|B)$ is given by
\begin{align}
l\leq H_{\mathrm{min}}(A|B).
\end{align}
However, this claim is restricted to the extraction of perfectly uniform randomness. We take the smooth min-entropy to extend this concept to a general case of approximately uniform randomness.
Roughly, for any $\epsilon\geq0$, the $\epsilon$-smooth min-entropy of $A$ given $B$, denoted $H^{\epsilon}_{\mathrm{min}}(A|B)$, is described as the maximum value of $H_{\mathrm{min}}(A|B)$ evaluated for all
density operators $\tilde{\rho}$ that are $\epsilon$-close to $\rho$ in terms of the purified distance.

\begin{defi}
Let $\epsilon\geq0$ and $\rho_{AB}\in S_{\leq}(\mathcal{H}_{AB})$. The min-entropy of $A$ conditioned on $B$ is given by
\begin{align}
H_{\mathrm{min}}(A|B)_\rho:=\mathop{\mathrm{max}}_{\sigma_B\in S_=(H_B)}\mathrm{sup}\{\lambda\in R:\rho_{AB}\leq2^{-\lambda}\mathbb{I}_A\otimes\sigma_B\}.
\end{align}
Similarly, the $smooth$ $min-entropy$ of $A$ conditioned on $B$ is given by
\begin{align}
H^\epsilon_{\mathrm{min}}(A|B)_\rho:=\mathop{\mathrm{max}}_{\rho_{AB}\in\mathcal{B}_{\rho_{AB}}^\epsilon}H_\mathrm{min}(A|B)_{\tilde{\rho}}.
\end{align}
Meanwhile, the $\epsilon$-ball of states close to $\rho\in\mathcal{S}_{\leq}(\mathcal{H})$ is given by
\begin{align}
\mathcal{B}^\epsilon(\rho):=\{\tilde{\rho}\in\mathcal{S}_{\leq}(\mathcal{H}):P(\rho,\tilde{\rho})\leq\epsilon\}.
\end{align}
\end{defi}

The quantum AEP is the statement that, in the limit of a large number of identical repetitions of a random quantum experiment, the output sequence is virtually certain to come from
the typical set, each member of which is almost equally likely. We take the conditional state that has the form
\begin{align}
\tau^{\otimes n}=p^{-1}\Pi\rho^{\otimes n}\Pi,
\end{align}
where $\Pi$ is a projector operator and $p=\mathrm{tr}(\Pi\rho^{\otimes n}\Pi)$ is the probability of successful error correction. We can bound the smooth min-entropy from a
conditional state $\tau^{\otimes n}$ to an i.i.d state $\rho^{\otimes n}$, i.e.,
\begin{align}
H^\epsilon_{\mathrm{min}}(A^n|B^n)_\tau=&\mathop{\mathrm{max}}_{\widetilde{\tau}} H^\epsilon_{\mathrm{min}}(A^n|B^n)_{\widetilde{\tau}}\\
=&\mathop{\mathrm{max}}_{\widetilde{\tau},\sigma_B}-\log\|\sigma^{\frac{1}{2}}_B\tilde{\tau}^{\otimes n}_{AB}\sigma^{\frac{1}{2}}_B\|_\infty\\
=&\mathop{\mathrm{max}}_{\widetilde{\tau},\sigma_B}-\log\|\sigma^{\frac{1}{2}}_B\tilde{p}^{-1}\Pi\tilde{\rho}^{\otimes n}_{AB}\Pi\sigma^{\frac{1}{2}}_B\|_\infty\\
\geq&\mathop{\mathrm{max}}_{\widetilde{\tau},\sigma_B}-\log\|\sigma^{\frac{1}{2}}_B\tilde{\rho}^{\otimes n}_{AB}\sigma^{\frac{1}{2}}_B\|_\infty-\log\frac{1}{p}\\
=&\mathop{\mathrm{max}}_{\widetilde{\rho}} H^\epsilon_{\mathrm{min}}(A^n|B^n)_{\widetilde{\rho}}-\log\frac{1}{p}\\
=&H^\epsilon_{\mathrm{min}}(A^n|B^n)_\rho-\log\frac{1}{p}
\end{align}
The smooth relative min-entropy can be related to relative R$\acute{e}$nyi entropies given by
\begin{align}\label{Renyi}
H^\epsilon_{\mathrm{min}}(A^n|B^n)_\rho\geq H_\alpha(A^n|B^n)_\rho-\frac{g(\epsilon)}{\alpha-1},  \end{align} where $\alpha\in(1,2]$ and $g(\epsilon)=-\log{1-\sqrt{1-\epsilon^2}}$.
For $\alpha=1+\frac{1}{\sqrt{n}}$, we obtain
\begin{align}
H^\epsilon_{\mathrm{min}}(A^n|B^n)_\rho\geq H_\alpha(A^n|B^n)_\rho-\sqrt{n}\log\frac{2}{\epsilon^2}.
\end{align}
According to Lemma $6.3$ in Ref.\cite{Tomamichel2012A}, we have $\alpha$-R$\acute{e}$nyi entropy by using $von$ $Neumann$ $entropy$ given by
\begin{align}
H_\alpha(A|B)_\rho=H(A|B)_\rho-4(\alpha-1)(\log\upsilon)^2,
\end{align}
where $\upsilon=\sqrt{2^{-H_{3/2}(A|B)_\rho}}+\sqrt{2^{H_{1/2}(A|B)_\rho}}+1$. In the four-state DM-CVQKD protocol, the convergence parameter becomes
$\upsilon\leq\sqrt{2^{-H_{\mathrm{min}}(A|B)_\rho}}+\sqrt{2^{H_{\mathrm{max}}(A|B)_\rho}}+1\leq4$.
Up to now, we can achieve
\begin{align}\label{tau-rho}
H^\epsilon_{\mathrm{min}}(A^n|B^n)_\tau\geq H(A^n|B^n)_\rho-16\sqrt{n}-\sqrt{n}\log\frac{2}{\epsilon^2}-\log\frac{1}{p}.
\end{align}

In order to establish the relationship between $H^\epsilon_{\mathrm{min}}(A^n|B^n)_{\tau}$ and $H(A^n|B^n)_{\tau}$, we employ the duality property of the conditional
von Neumann entropy. Assume $\rho_{ABC}$ is a purification of $\rho_{AB}$ and $\tau_{A^nB^nC^n}=\frac{1}{p}(\Pi\otimes\mathbb{I}_C)\rho^{\otimes n}_{ABC}(\Pi\otimes\mathbb{I}_C)$ is that of $\tau_{A^nB^n}$. Applying Theorem $9$ from \cite{tomamichel2009a} to $\rho^{\otimes n}$, we obtain
\begin{align}
\frac{1}{n}H^\epsilon_{\mathrm{min}}(A^n|C^n)_\rho\geq H(A|C)_\rho-\frac{8}{\sqrt{n}}\sqrt{\log{\frac{2}{\epsilon^2}}}.
\end{align}
According to the definition of the smooth min-entropy, there exists an operator
$\bar{\rho}_{A^nC^n}$ satisfying the properties
\begin{align}
\lambda\cdot(\mathbb{I}_A\otimes\rho_C)^{\otimes n}-\bar{\rho}_{A^nC^n}\geq0, \quad
\|\rho^{\otimes n}_{AC}-\bar{\rho}_{A^nC^n}\|_1\leq\epsilon,
\end{align} where $\lambda:=2^{-H(A^n|C^n)_\rho+8\sqrt{n\log{\frac{2}{\epsilon^2}}}}$.
Similarly, we define $\bar{\tau}_{A^nB^nC^n}:=\frac{1}{\bar{p}}\Pi\bar{\rho}_{A^nB^nC^n}\Pi$, and due to $\Pi<\mathbb{I}_{ABC}$,  we have
\begin{align}\label{trace-norm}
&T=\frac{1}{2}\|\tau_{A^nC^n}-\bar{\tau}_{A^nC^n}\|_1\leq\frac{\epsilon}{p}
\\
&\bar{\tau}_{A^nC^n}\leq\frac{1}{\bar{p}}\bar{\rho}_{A^nC^n}\leq\frac{1}{\bar{p}}\lambda\cdot(\mathrm{id}_A\otimes\rho_E)^{\otimes n}.
\end{align}
In Eq.(\ref{trace-norm}), we use the assumption that $\frac{p}{2}\leq\bar{p}\leq2p$. The conditional von Neumann entropy can be described as
\begin{align}
H(A^n|C^n)_{\bar{\tau}}&=\mathrm{tr}(\bar{\tau}_{A^nC^n}(\mathbb{I}_A\otimes\log\sigma^{\otimes n}_C-\log\bar{\tau}_{A^nC^n}))-\mathrm{tr}(\bar{\tau}_{C^n}(\log\sigma^{\otimes n}_C-\log\bar{\tau}_{C^n}))\\
&\geq\mathrm{tr}(\bar{\tau}_{A^nC^n}(\mathbb{I}_A\otimes\log\sigma^{\otimes n}_C-\log\bar{\tau}_{A^nC^n}))\\
&\geq\mathrm{tr}(\bar{\tau}_{A^nC^n}(\mathbb{I}_A\otimes\log\sigma^{\otimes n}_C-\log(\mathbb{I}_A\otimes\rho_E)^{\otimes n}-\log\lambda-\log\frac{1}{\bar{p}}))\\
&\geq\mathrm{tr}(\bar{\tau}_{A^nC^n}(-\log\lambda-\log\frac{1}{\bar{p}}))\\
&\geq H(A^n|C^n)_\rho-8\sqrt{n\log\frac{2}{\epsilon^2}}-\log\frac{2}{p}
\end{align}
As stated previously that $\rho_{ABC}$ is a purification of $\rho_{AB}$, which implies $H(A|C)=-H(A|B)$, we have the inequations
\begin{align}
&-H(A^n|B^n)_{\bar{\tau}}\geq-H(A^n|B^n)_\rho-8\sqrt{n\log\frac{2}{\epsilon^2}}-\log\frac{2}{p},
\\ \label{eq113}
&H(A^n|B^n)_\rho\geq H(A^n|B^n)_{\bar{\tau}}-8\sqrt{n\log\frac{2}{\epsilon^2}}-\log\frac{2}{p}.
\end{align}
Based on Theorem 1 in Ref.\cite{audenaert2007a}, we have
\begin{align}
H(A^n|B^n)_{\bar{\tau}}&\geq H(A^n|B^n)_{\tau}-T\log(d-1)-H((T,1-T))\\
&\geq H(A^n|B^n)_{\tau}-4\frac{\epsilon}{p}+\frac{\epsilon}{p}\log\frac{\epsilon}{p}+(1-\frac{\epsilon}{p})\log(1-\frac{\epsilon}{p})\\
&\geq H(A^n|B^n)_{\tau}-4\frac{\epsilon}{p}
\end{align}
where the last inequality holds because of $0<\frac{\epsilon}{p}<1$. Combining with Eq.(\ref{eq113}), it becomes
\begin{align}
H(A^n|B^n)_\rho\geq H(A^n|B^n)_\tau-4\frac{\epsilon}{p}-8\sqrt{n\log\frac{2}{\epsilon^2}}-\log\frac{2}{p}.
\end{align}
According to Eq. (\ref{tau-rho}), we obtain
\begin{align}
H^\epsilon_{\mathrm{min}}(A^n|B^n)_{\tau}\geq H(A^n|B^n)_{\tau}-\Delta_{AEP},
\end{align}
with
$
\Delta_{AEP}=\sqrt{n}(16+\log\frac{2}{\epsilon_\mathrm{sm}^2}+8\sqrt{\log\frac{2}{\epsilon_\mathrm{sm}^2}})+4\frac{\epsilon_\mathrm{sm}}{p}+\log\frac{2}{p^2}.
$

In the DM-CVQKD protocol, the above-mentioned random variables are subject to the independent and identical distribution and have the same probability falling to the four different quadrants.
Unfortunately, the practical entropy $H(U)=-\sum^4_{i=1}p_i\log{p_i}$ may be less than that of the ideal one for the finite case. In what follows, we give the lower bound of the practical entropy.
Let the maximum likelihood estimator (MLE) of $H(U)$ be
\begin{align}
\hat{H}_{\mathrm{MLE}}(U):=-\sum^4_{i=1}\hat{p}_i\log\hat{p}_i,
\end{align}
where $p_i$ is the probability of data falling into the $i$th quadrant. Then $\hat{H}_\mathrm{MLE}(U)$ is negatively biased everywhere
\begin{align}
\mathbb{E}_p\hat{H}_\mathrm{MLE}(U)\leq H(U).
\end{align}
where $\mathbb{E}_p$ denotes the conditional expectation for the given $p$.
The empiric is bounded on $H(U)$ but can not be observed in experiments.
According to the derived bound in Theorem (Antos and Kontoyiannis, 2001)~\cite{antos2001convergence},\begin{align}
\mathrm{Pr}[|\hat{H}_\mathrm{MLE}-\mathbb{E}\hat{H}_\mathrm{MLE}|\geq\epsilon]\leq2\exp{(-n\epsilon^2/2\log^2_2n)},
\end{align}
where $n$ is the number of point that the distribution concentrates on, we obtain a lower bound of $\hat{H}(U)$ due to the measurable $\hat{H}_\mathrm{MLE}$. Namely, we have
\begin{align}
n\hat{H}(U)\geq n\hat{H}_\mathrm{MLE}-\Delta_n\log n\sqrt{2\log\frac{2}{\epsilon_\mathrm{ent}}},
\end{align}
which holds with a probability of at least $1-\epsilon$. In the two-state DM-CVQKD protocol, $n$ equals to $2$ and it becomes $4$ in the four-state DM-CVQKD protocol in a similar manner.

\section{Gaussian De Finetti Reduction}
\label{de finetti reduction}

Alice prepares one of four coherent states that are described in the Hilbert space. Let $H_A$ and $H_B$ be two Hilbert spaces of dimension $n$, constituting the $4n$-dimensional Hilbert space $H_{2,2,n}:=H_A\oplus H_B\oplus H_A'\oplus H_B'\cong \mathbb{C}^{4n}$.
The Fock space $F_{2,2,n}$ associated with $H_{2,2,n}$ is the infinite-dimensional Hilbert space
\begin{align}
F_{2,2,n}:=\bigoplus^\infty_{k=1}\mathrm{Sym}^k(H_{2,2,n}),
\end{align}
where $\mathrm{Sym}^k$ stands for the symmetric part of $H^{\otimes k}$.

Performing any unitary $u\in U(n)$ on these variables, we have
\begin{align}
z_i\rightarrow uz_i, \quad\forall i\in\{1,2\}, \quad \mathrm{and} \quad z'_j\rightarrow \bar{u}z'_j,  \quad\forall j\in\{1,2\},
\end{align}
where $\bar{u}$ denotes the complex conjugate of the unitary matrix $u$. The switch between Segal-Bargmann representation and representation in terms of annihilation and creation
operators can be expressed as
\begin{align}
z_{k,1}\leftrightarrow a^\dag_k, \quad z_{k,2}\leftrightarrow b'^\dag_k, \quad
z'_{k,1}\leftrightarrow b^\dag_k ,\quad z'_{k,2}\leftrightarrow a'^\dag_k.
\end{align}
For the convenience of characterizing the coherent states, we take the four operators $Z_{11},Z_{12},Z_{21},Z_{22}$ as follows
\begin{align}
Z_{11}=\sum^\infty_{k=1}z_{k,1}z'_{k,1}, Z_{12}=\sum^\infty_{k=1}z_{k,1}z'_{k,2},
Z_{21}=\sum^\infty_{k=1}z_{k,2}z'_{k,1},  Z_{22}=\sum^\infty_{k=2}z_{k,1}z'_{k,2}.
\end{align}

\begin{defi}
(Symmetric subspace). For an integer $n\geq1$, the symmetric subspace $F^{U(n)}_{2,2,n}$ is the subspace of functions $\psi\in F_{2,2,n}$ such that
\begin{align}
W_u\psi=\psi,
\end{align} with $\forall u\in U(n)$.
\end{defi}

The definition of $\psi$ can be found in Ref.\cite{Leverrier2016}.
The domain of the factor space is given by
\begin{align}
\mathcal{D}:=\{\Lambda\in M_2(\mathbb{C}):\mathbb{I}_p-\Lambda\Lambda^\dag\geq0\},
\end{align} and the cut-off versions $D_\eta$ is given by
\begin{align}
\mathcal{D}_\eta:=\{\Lambda\in M_2(\mathbb{C}):\eta\mathbb{I}_p-\Lambda\Lambda^\dag\geq0\},
\end{align}
for $\eta\in[0,1]$.

\begin{defi}
(Generalized coherent states). For $n\geq1$, the coherent state $\psi_{\Lambda,n}$ with $\Lambda\in\mathcal{D}$ is given by
\begin{align}
\psi_{\Lambda,n}=(Z_{1,1},\cdots,Z_{2,2})=\det(1-\Lambda\Lambda^\dag)^{n/2}\det\exp(\Lambda^TZ),
\end{align}
where $\mathrm{Z}$ is the $2\times2$ matrix $[Z_{i,j}]_{i,j\in\{1,2\}}$.
\end{defi}

Subsequently, we show the modified protocol from an initial protocol $\mathcal{E}_0$ known to be secure against Gaussian collective attacks, by prepending
an energy test and an additional privacy amplification test. The CPTP map, which is the CVQKD protocol, transforming the infinite-dimensional Hilbert
space $(\mathcal{H}_A\otimes\mathcal{H}_B)^{\otimes n}$ to the sets of pairs $(S_A,S_B)$ of $l$-bit strings and $C$, a transcript of the classical communication.
A method to characterize a given CPTP map $\mathcal{E}$ is to compare it to an idealized CPTP map $\mathcal{F}$, which can be generated by concatenating the protocol $\mathcal{E}$ with a map
$\mathcal{S}$ transforming $(S_A,S_B)$ to two strings of perfect key, that is $\mathcal{F}=\mathcal{S}\circ\mathcal{E}$.
An operational way of quantifying the security is to bound the diamond distance between the two maps \cite{PhysRevLett.110.030502} \begin{align}
\|\mathcal{E}-\mathcal{F}\|_\diamond:=\sup_{\rho_{ABE}}\|(\mathcal{E}-\mathcal{F})\otimes\mathrm{id}_\mathcal{K}(\rho_{ABE})\|_1,
\end{align}
where the supremum is taken over density operators on $(\mathcal{H}_A\otimes\mathcal{H}_B)^{\otimes n}\otimes\mathcal{K}$ for any auxiliary system $\mathcal{K}$.

We will give a reduction of the security against general attacks to that against Gaussian collective attacks, for which security has already been proved in previous sections.
Let us suppose that the DM-CVQKD protocol, $\mathcal{E}_0$, is secure against Gaussian collective attacks. In order to ensure the success of reduction, we perform an energy test, $\mathcal{T}$,
a CP map taking a state in a slightly larger Hilbert space, $(\mathcal{H}_A\otimes\mathcal{H}_B)^{\otimes(n+k)}$, applying a random unitary $u\in U(n+k)$, measuring the last $k$ modes
and comparing the outcome to a threshold fixed in advance. The test is passed if the measurement result is less than the threshold, and then the global state is restricted to state with low energy.
We apply another CP map $\mathcal{P}$ which projects a state on $F_{1,1,n}=(\mathcal{H}_A\otimes\mathcal{H}_B)^{\otimes n}$ to a low-dimensional Hilbert space $F^{\leq K}_{1,1,n}$
with less than $K$ photons overall in the $2n$ modes shared by Alice and Bob.

Let $\mathcal{E}_0$ be a CVQKD protocol, which is $\epsilon$-secure against Gaussian collective attacks~\cite{PhysRevLett.102.020504}, i.e.,
\begin{align}\label{collective-attack}
\|((\mathcal{E}_0-\mathcal{F}_0)\otimes\mathbb{I})(|\Lambda,n\rangle\langle\Lambda,n|)\|_1\leq\epsilon,
\end{align}
where $|\Lambda,n\rangle$ is the generalized coherent state and $\mathcal{F}_0:=\mathcal{S}\circ\mathcal{E}_0$ is a ideal version under Gaussian collective attacks.
We have the following definitions
\begin{align}
&\mathcal{T}:\mathfrak{B}(F_{1,1,n+k})\rightarrow \mathfrak{B}(F_{1,1,n})\otimes\{\mathrm{passes/aborts}\},\\
&\mathcal{P}:\mathfrak{B}(F_{1,1,n})\rightarrow \mathfrak{B}(F^{\leq K}_{1,1,n}),\\
&\mathcal{R}:\{0,1\}^l\times\{0,1\}^l\rightarrow\{0,1\}^{l'}\times\{0,1\}^{l'}.
\end{align}  $\mathcal{T}$ is an energy test that applies the Haar measurement on $U(n+k)$, measure the last $k$ modes of $A$ and $B$, and check whether the measurement outputs pass the tests.
The measurement results pass the test if Alice's measurement result $\alpha_1,\cdots,\alpha_n$ and Bob's measurement result $\beta_1,\cdots,\beta_k$ satisfy the constraints $\Sigma^n_{i=1}|\alpha_i|^2\leq nd_A$
and $\Sigma^k_{i=1}|\beta_i|^2\leq kd_B$. The first $n$ modes will be kept if they pass the test, and will be aborted otherwise.
$\mathcal{P}$ is a dimension reduction that maps state $\rho\in\mathfrak{B}(F_{1,1,n})$ to $\Pi_{\leq k}\rho\Pi_{\leq k}\in\mathfrak{B}(F^{\le K}_{1,1,n})$. It ensures the state in protocol $\mathcal{E}_0$
in a finite-dimensional subspace.
$\mathcal{R}$ is a privacy amplification procedure that inputs two $l$-bit strings and outputs $l'$-bit strings.

Finally, we obtain the DM-CVQKD protocol as follows
\begin{align}
\mathcal{E}=\mathcal{R}\circ\mathcal{E}_0\circ\mathcal{T}.
\end{align}

For the security analysis of the protocol $\mathcal{E}$ against arbitrary attacks, one needs to bound $\|\mathcal{E}-\mathcal{F}\|_\diamond$. Therefore, we have to show the finite energy version of de Finetti theorem\cite{PhysRevLett.118.200501}. In the CVQKD protocol, these states are prepared in infinite-dimensional Hilbert space. We give a dimension
reduction while the energy is bounded.
For the dimension of $F^{U(n),\leq K}_{2,2,n}$, we follow the results derived in Ref.~\cite{PhysRevLett.118.200501}, which gives $\dim{F^{U(n),\leq K}_{2,2,n}}=\binom{K+4}{4}$.

\begin{lemma}\label{finite-energy-version}
For $n\geq5$ and $\eta\in[0,1]$, if $K\leq\frac{\eta}{1-\eta}(n-5)$, then the following operator inequality holds
\begin{align}
\int_{\mathcal{D}_\eta}|\Lambda,n\rangle\langle\Lambda,n|d\nu\mu_n(\Lambda)\geq(1-\varepsilon)\Pi_{\leq K}
\end{align}
with $\Pi_{\leq K}:=\sum^K_{k=0}\Pi_{=k}$,
$
\varepsilon\leq\frac{2(N+K)^7}{N^3}\exp(-\frac{2N^3}{(N+K)^2\ln2})
$ and $N=n-5$.
\end{lemma}

\begin{theo}
For $K\geq\frac{n}{1-\eta}$, it holds that
\begin{align}
T(n,\eta):=\mathrm{tr}\int_{\mathcal{D}_\eta}|\Lambda,n\rangle\langle\Lambda,n|d\mu_n(\Lambda)\leq\frac{K^4}{12}.
\end{align}
\end{theo}
\begin{proof}
The trace of the volume on $\mathcal{D}_{\eta}$ is given by
\begin{align}
\mathrm{tr}\int_{\mathcal{D}_\eta}|\Lambda,n\rangle\langle\Lambda,n|d\mu_n(\Lambda)&=\int^\eta_0\int^\eta_0q(x,y)\mathrm{d}x\mathrm{d}y\\
&=\frac{(n-1)(n-2)^2(n-3)}{2}\int^\eta_0\int^\eta_0\frac{(x-y)^2}{(1-x)^4(1-y)^4}\mathrm{d}x\mathrm{d}y\\
&=\frac{(n-1)(n-2)^2(n-3)}{12(1-\eta)^4}\\
&\leq\frac{n^4}{12(1-\eta)^4}\\
&=\frac{K^4}{12}.
\end{align}
The last equation is achieved with $K=\frac{n}{1-\eta}$.
\end{proof}

To bound the diamond distance of $\|\mathcal{E}-\mathcal{F}\|_\diamond$ \cite{PhysRevLett.118.200501}, it can be described as
\begin{align}
\|\mathcal{E}-\mathcal{F}\|_\diamond&\leq\|\tilde{\mathcal{E}}\circ\mathcal{T}-\tilde{\mathcal{F}}\circ\mathcal{T}\|_\diamond+\|\mathcal{E}-\tilde{\mathcal{E}}\circ\mathcal{T}\|_\diamond+\|\mathcal{F}-\tilde{\mathcal{F}}\circ\mathcal{T}\|_\diamond\\
&\leq\|(\tilde{\mathcal{E}}-\tilde{\mathcal{F}})\circ\mathcal{T}\|_\diamond+\|\mathcal{R}\circ\mathcal{E}_0(\mathrm{id}-\mathcal{P})\circ\mathcal{T}\|_\diamond+\|\mathcal{S}\circ\mathcal{R}\circ\mathcal{E}_0\circ(\mathrm{id}-\mathcal{P})\mathcal{T}\|_\diamond\\
&\leq\|\tilde{\mathcal{E}}-\tilde{\mathcal{F}}\|_\diamond+2\|(\mathrm{id}-P)\circ\mathcal{T}\|_\diamond
\end{align}
where $\tilde{\mathcal{E}}=\mathcal{R}\circ\mathcal{E}_0\circ\mathcal{P}$ and $\tilde{\mathcal{F}}=\mathcal{S}\circ\tilde{\mathcal{E}}$. We note the CP maps $\mathcal{T}$, $\mathcal{S}$, $\mathcal{R}$ and $\mathcal{E}_0$ would not increase the diamond norm.
Taking $\tilde{\mathcal{E}}=\overline{\mathcal{E}}\circ\mathcal{P}$ with $\mathcal{P}$ being the trace non-increasing CP map, the diamond norm becomes
\begin{align}\label{diamond norm}
\|\mathcal{E}-\mathcal{F}\|_\diamond\leq\|\overline{\mathcal{E}}-\overline{\mathcal{F}}\|_\diamond+2\|(\mathbb{I}-P)\circ\mathcal{T}\|_\diamond.
\end{align}
For the first part on the right of the inequality, we bound the diamond distance as follows.

\begin{theo}
If $\mathcal{E}_0$ is $\varepsilon$-secure against Gaussian collective attacks, then
\begin{align}
\|\overline{\mathcal{E}}-\overline{\mathcal{F}}\|_\diamond=\|\mathcal{R}\circ(\mathcal{E}_0-\mathcal{F}_0)\circ\mathcal{P}^{\leq K}\|_\diamond\leq2T(n,\eta)\varepsilon.
\end{align}
where $T(n,\eta)=\frac{(n-1)(n-2)^2(n-3)}{12(1-\eta)^4}$ and $\overline{\mathcal{E}}=\mathcal{R}\circ\mathcal{E}_0\circ\mathcal{P}^{\leq K}$.
\end{theo}

\begin{proof}
To restrict $\varepsilon$ in Lemma~\ref{finite-energy-version} less than $\frac{1}{2}$,  for $\alpha=\frac{K}{N}$, we have a map $N^*:= [1, \infty]\rightarrow N$ such that
\begin{align}
N^*(\alpha)=\max{\{38,\min\{N\in\mathbb{N}:2(1+\alpha)^7N^4\exp{(-\frac{2N}{(1+\alpha)^2\ln2})\leq\frac{1}{2}}}\}\}.
\end{align}
We also define the positive operator
\begin{align}
P_\eta:=\int_{\mathcal{D}_\eta}|\Lambda,n\rangle\langle\Lambda,n|\mathrm{d}\nu_n(\Lambda),
\end{align}
where $d\nu_n(\Lambda)=\frac{(n-1)(n-2)^2(n-3)}{\pi^4\det(\mathbb{I}_2-\Lambda\Lambda^\dag)^4}\mathrm{d}\lambda_{11}\mathrm{d}\lambda_{12}\mathrm{d}\lambda_{21}\mathrm{d}\lambda_{22}$.
Let $\tau^\eta_\mathcal{H}$ be the normalized state of the projector $P_\eta$ given by
\begin{align}
\tau^\eta_\mathcal{H}=\mathrm{tr}(P_\eta)P_\eta.
\end{align}
We obtain $|\Phi^\eta\rangle=(\sqrt{\tau^\eta_\mathcal{H}}\otimes\mathbb{I})|\Phi\rangle$, a purification of $\tau^\eta_\mathcal{H}$, where $|\Phi\rangle$ can be expressed by the orthonormal basis of $F^{U(n)}_{2,2,n}$,
$|\Phi\rangle_{\mathcal{H}\mathcal{N}}=\sum_i|\nu_i\rangle_\mathcal{H}\langle\nu_i|\mathcal{N}$.
Assume $\rho$ is an arbitrary density operator on $F^{U(n),\leq K}_{2,2,n}$ and $M=p\tau^{-1/2}\rho\tau^{-1/2}$ is non negative operator.
After performing the measurement $\mathcal{M}=\{M,\mathbb{I}-M\}$ on state $\Phi^\eta$, we obtain
\begin{eqnarray}
\begin{split}
\tau^\eta_\mathcal{H}&=\mathrm{tr}_\mathcal{N}(|\Phi^\eta\rangle\langle\Phi^\eta|_{\mathcal{H}\mathcal{N}}) \\ \nonumber
&=\mathrm{tr}_\mathcal{N}((1\otimes M^{1/2})|\Phi^\eta\rangle\langle\Phi^\eta|(1\otimes M^{1/2})) \\ \nonumber
&=\mathrm{tr}((\tau^{1/2}\otimes M^{1/2})\sum_{i,j}|\nu_i\rangle\langle\nu_j|\otimes|\nu_i\rangle\langle\nu_j|(\tau^{1/2}\otimes M^{1/2}))\\ \nonumber
&=\tau_{1/2}p\tau_{-1/2}\rho\tau_{-1/2}\tau_{1/2}\\ \nonumber
&=p\rho. \nonumber
\end{split}
\end{eqnarray}
Then we have
\begin{align}
\rho_{\mathcal{H}\mathcal{N}}=2T(n,\eta)\cdot\tau_{\mathcal{H}\mathcal{N}}.
\end{align}
where $p=\frac{1}{2T(n,\eta)}$. The hypothesis that protocol $\mathcal{E}_0$ is $\epsilon$-secure against collective attacks  in Eq.~(\ref{collective-attack}) implies
\begin{align}
\|((\mathcal{E}_0-\mathcal{F}_0)\otimes\mathbb{I})(|\Lambda,n\rangle\langle\Lambda,n|)\|_1\leq\epsilon
\end{align}
Let $\Delta$ be a linear map from End$(F^{\leq K}_{1,1,n})$ to End$(\mathcal{H}')$. Particularly, $\Delta$ may be the difference between the
CPTP maps $\overline{\mathcal{E}}$ and $\overline{\mathcal{F}}$, where End$(\mathcal{L})$ denotes the space of all endomorphisms on $\mathcal{L}$.
The linear CPTP map $K_\pi$ such that $\Delta\circ\pi=K_\pi\circ\Delta$ reduces the diamond distance to the linear norm~\cite{PhysRevLett.102.020504}
\begin{align}
\|\Delta\|_\diamond\leq2T(n,\eta)\|(\Delta\otimes\mathrm{id})\tau^\eta_{\mathcal{H}\mathcal{N}}\|_1.
\end{align}
Finally, we have
\begin{eqnarray}
\begin{split}
\|\overline{\mathcal{E}}-\overline{\mathcal{F}}\|_\diamond&\leq2T(n,\eta)\|((\overline{\mathcal{E}}-\overline{\mathcal{F}})\otimes\mathrm{id})\tau^\eta_{\mathcal{H}\mathcal{N}}\|_1  \\
&=2T(n,\eta)\|(\mathcal{R}\circ(\mathcal{E}_0-\mathcal{F}_0)\circ\mathcal{P}^{\leq K}\otimes\mathrm{id})\tau^\eta_{\mathcal{H}\mathcal{N}}\|_1  \\
&\leq2T(n,\eta)\|((\mathcal{E}_0-\mathcal{F}_0)\circ\mathcal{P}^{\leq K}\otimes\mathrm{id})\tau^\eta_\mathcal{H}\|_1  \\
&=2T(n,\eta)\|((\mathcal{E}_0-\mathcal{F}_0)\otimes\mathrm{id})(T(n,\eta)^{-1}\int_\mathcal{D}|\Lambda,n\rangle\langle\Lambda,n|\mathrm{d}\nu_n(\Lambda))\|_1  \\
&\leq2\epsilon\|\int_\mathcal{D}\mathrm{d}\nu_n(\Lambda)\|_1\\
&=2T(n,\eta)\epsilon,
\end{split}
\end{eqnarray}
which completes the proof of this theorem.
\end{proof}

Moreover, the diamond norm of second part on the right of the inequality of Eq.~(\ref{diamond norm}) can be restricted to be $\varepsilon$-small.
\begin{lemma}
For any $d_A,d_B\geq0$ and integer $K\leq n(d_A+d_B)$, owing to $U_n^d\leq2T^d_n$, we have
\begin{align}
\mathbb{I}_{\mathcal{H}^{\otimes n}_A\otimes\mathcal{H}^{\otimes n}_B}-\mathcal{P}^{\leq K}\leq2T^{d_A}_n\otimes\mathbb{I}_{H^{\otimes n}_B}+2\mathbb{I}_{H^{\otimes n}_A}\otimes T^{d_B}_n.
\end{align}
\end{lemma}

\begin{theo}
Let $\rho$ be a rotationally invariant state on $\mathcal{H}^{\otimes(n+k)}$. For any $d\geq0$, we have
\begin{align}
\mathrm{tr}[(T_n^{d'}\otimes(\mathbb{I}-T^d_k))\rho]\leq\varepsilon
\end{align}
where $d'=g(n,k,\varepsilon)d$ with
$
g(n,k,\varepsilon)=\frac{1+2\sqrt{\frac{\ln(2/\varepsilon)}{n}}+\frac{2\ln(2/\varepsilon)}{n}}{1-2\sqrt{\frac{\ln(2/\varepsilon)}{k}}}.
$
\end{theo}

\begin{proof}
Based on the definition of $T^d_n:=\frac{1}{\pi^n}\int_{\Sigma^n_{i=1}|\alpha_i|^2\geq nd}|\alpha_1\rangle\langle\alpha_1|\otimes\cdots\otimes|\alpha_n\rangle\langle\alpha_n|\mathrm{d}\alpha_1\cdots\alpha_n$, $\mathrm{tr}[T^d_n\rho]$ is the probability of applying measurement on $n$ modes of $\rho$ with outcomes satisfying $\Sigma^k_{i=1}|\alpha_i|^2\geq nd$.
In the DM-CVQKD protocol, the  prepared coherent states are $|\alpha e^{i(2k+1)\pi/4}\rangle$ and we take $\alpha_i=\sqrt{x^2_i+p^2_i}$ for the security analysis.
In the energy test, $n+k$ modes of state $\rho$ are measured with outcomes $\alpha_1,\cdots,\alpha_{n+k}$, where the first $n$ modes will be kept if the
last $k$ modes meet the constraint $\Sigma^k_{i=1}|\alpha_i|^2\leq nd$. $\mathrm{Tr}[((T^{d'}_n\otimes(1-T^d_k))\rho]$ is the probability that the energy test does not work with
\begin{align}
Y_n:=\Sigma^n_{i=1}|\alpha_i|^2\geq nd' \quad \mathrm{and} \quad Y_k:=\Sigma^k_{i=1}|\alpha_{n+i}|^2\leq kd.
\end{align}
Then $\mathrm{tr}[(T_n^{d'}\otimes(\mathbb{I}-T^d_k))\rho]$ can be expressed as
\begin{align}
\mathrm{tr}[(T_n^{d'}\otimes(\mathbb{I}-T^d_k))\rho]&=\mathrm{Pr}[(Y_n\geq nd')\wedge(Y_k\leq kd)]\\
&\leq\mathrm{Pr}[kdY_n\geq nd'Y_k]
\end{align}
where the last inequality holds because the parameters are all positive. Because the amplitude $\alpha$ is subject to Gaussian distribution with variance $V_\alpha$,
$Y_n/V_\alpha$ and $Y_k/V_\alpha$ both obey the $\chi^2$ distribution with $Y_n/V_\alpha\thicksim\chi^2(n)$ and $Y_k/V_\alpha\thicksim\chi^2(k)$.
Consequently, we obtain the probability
\begin{align}
\mathrm{Pr}[kdY_n\geq nd'Y_k]&=\mathrm{Pr}[kdY_n/V_\alpha\geq nd'Y_k/V_\alpha]\\
&\leq\mathrm{Pr}[Y_n/V_\alpha\geq tnd']+\mathrm{Pr}[Y_k/V_\alpha\leq tkd],
\end{align} where $t$ is chosen such that
$
tkd=k-2\sqrt{k\ln\frac{2}{\varepsilon}}
$.
According to Lemma.\ref{laurent-massart}, the parameter $d'$ is chosen to meet the constraints \begin{align}
\mathrm{Pr}[Y_n/V_\alpha\geq tnd']\leq\frac{\varepsilon}{2}, \quad \quad \mathrm{Pr}[Y_k/V_\alpha\leq tkd]\leq\frac{\varepsilon}{2}.
\end{align}
Therefore, we obtains
\begin{align}
d'=\frac{1+2\sqrt{\frac{\ln(2/\varepsilon)}{n}}+\frac{2\ln(2/\varepsilon)}{n}}{1-2\sqrt{\frac{\ln(2/\varepsilon)}{k}}}d,
\end{align}
which completes the proof.
\end{proof}

\begin{lemma}
\label{enegy test}
For integers $n, k, d_A, d_B\geq1$, taking $K=n(d'_A+d'_B)$ for $d_{A/B}:=g(n,k,\varepsilon/4)d_{A/B}$, we have
\begin{align}
\|(\mathbb{I}-\mathcal{P}(n,K))\circ\mathcal{T}(k,d_A,d_B)\|_\diamond\leq\varepsilon
\end{align}
\end{lemma}
\begin{proof}
\begin{align}
\|(\mathbb{I}-\mathcal{P})\circ\mathcal{T}\|_\diamond&=\mathop{\mathrm{max}}_{\rho\in\mathcal{H}^{\otimes2(n+k)}_{AB}}\|(((\mathbb{I}-\mathcal{P})\circ\mathcal{T})\otimes\mathbb{I}_{H^{\otimes(n+k)}_{AB}})(\rho)\|_1\\
&\leq\mathop{\mathrm{max}}_{\rho\in\mathrm{Inv}(\mathfrak{S}(\mathcal{H}^{\otimes(n+k)}_{AB}))}\|(\mathbb{I}-\mathcal{P}^{\leq K})\circ\mathcal{T}\|_1\\
&\leq\mathop{\mathrm{max}}_{\rho\in\mathrm{Inv}(\mathfrak{S}(\mathcal{H}^{\otimes(n+k)}_{AB}))}\|(U^{d'_A}_n\otimes\mathbb{I}+\mathbb{I}\otimes U^{d'_B}_n)\circ((\mathbb{I}-T^{d_A}_k)\otimes(T^{d_B}_k-\mathbb{I}))(\rho)\|\\
&\leq\mathop{\mathrm{max}}_{\rho\in\mathrm{Inv}(\mathfrak{S}(\mathcal{H}^{\otimes(n+k)}_{A}))}\|(U^{d'_A}_n\circ\mathbb{I}-T^{d_A}_k)(\rho)\|_1+
\mathop{\mathrm{max}}_{\rho\in\mathrm{Inv}(\mathfrak{S}(\mathcal{H}^{\otimes(n+k)}_{B}))}\|(U^{d'_B}_n\circ(\mathbb{I}-T^{d_B}_k))(\rho)\|\\
&\leq2\mathop{\mathrm{max}}_{\rho\in\mathrm{Inv}(\mathfrak{S}(\mathcal{H}^{\otimes(n+k)}_{A}))}\|(T^{d'_A}_n\circ(\mathbb{I}-T^{d_A}_k))(\rho)\|_1
+2\mathop{\mathrm{max}}_{\rho\in\mathrm{Inv}(\mathfrak{S}(\mathcal{H}^{\otimes(n+k)}_{B}))}\|(T^{d'_B}_n\circ(\mathbb{I}-T^{d_B}_k))(\rho)\|_1\\
&\leq\varepsilon
\end{align}
\end{proof}

\begin{theo}
If the protocol $\mathcal{E}_0$ is $\varepsilon$-secure against Gaussian collective attacks, then the protocol $\mathcal{E}=\mathcal{R}\circ\mathcal{E}_0\circ\mathcal{P}$ is
$\varepsilon'$-secure against general attacks with
\begin{align}
\varepsilon'\leq\frac{K^4}{6}\varepsilon+2\varepsilon.
\end{align}
\end{theo}

\bibliographystyle{apsrev4-1}
\bibliography{Supplementaryref}